\documentclass[prb,a4,
              twocolumn,
              showpacs,amsmath,amssymb]{revtex4-1}

\usepackage{graphicx}
\usepackage{graphicx,subfigure,epsfig}
\usepackage{color}
\usepackage{soul}
\usepackage{float}
\usepackage{amsmath}
\usepackage{amssymb}
\usepackage{ifthen}
\usepackage{alltt}
\usepackage{fancyhdr}
\usepackage{verbatim}
\usepackage{moreverb}
\usepackage{colortbl}
\usepackage{multirow}
\usepackage{bigstrut}
\usepackage{wrapfig}
\usepackage{enumerate}
\usepackage{rotating}
\usepackage{epstopdf}

\def\=#1{\overline{\overline #1}}
\def\e{\begin{equation}}
\def\f{\end{equation}}
\def\_#1{{\bf #1}}

\def\.{\cdot}

\def\r#1{(\ref{#1})}
\def\=#1{\overline{\overline{#1}}}

\def\aeeo{\alpha_{\rm ee}^{\rm co}}
\def\aeer{\alpha_{\rm ee}^{\rm cr}}
\def\aemo{\alpha_{\rm em}^{\rm co}}
\def\aemr{\alpha_{\rm em}^{\rm cr}}
\def\ameo{\alpha_{\rm me}^{\rm co}}
\def\amer{\alpha_{\rm me}^{\rm cr}}
\def\ammo{\alpha_{\rm mm}^{\rm co}}
\def\ammr{\alpha_{\rm mm}^{\rm cr}}

\def\=#1{\overline{\overline{#1}}}

\begin{document}

\title{Polarizabilities of nonreciprocal bianisotropic particles}
\author{M.~S.~Mirmoosa, Y.~Ra'di, V.~S.~Asadchy, C.~R.~Simovski, and S.~A.~Tretyakov}
\affiliation{Department of Radio Science and Engineering, School of Electrical Engineering, Aalto University, P.O. Box 13000, FI-00076 AALTO, Finland}
\date{\today }


\begin{abstract}
For two electrically small nonreciprocal scatterers an analytical electromagnetic model of polarizabilities is developed. Both particles are bianisotropic: the so-called Tellegen-omega particle and moving-chiral particle. Analytical results are compared to the full-wave numerical simulations. Both models satisfy to main physical restrictions and leave no doubts in the possibility to realize these particles experimentally. This paper is a necessary step towards applications of nonreciprocal bianisotropic particles such as perfect electromagnetic isolators, twist polarizers, thin-sheet phase shifters, and other devices.
\end{abstract}

\maketitle

\section{Introduction}

In terms of electromagnetic properties of materials, the most general linear material is the bianisotropic medium in which the relations between the four field vectors $\_E$, $\_H$, $\_D$, and $\_B$ are defined by four general dyadic coefficients, as in
\e
\left[ \begin{array}{c} \mathbf{D} \\ \mathbf{B}\end{array} \right]
=\left[ \begin{array}{cc} \overline{\overline{\epsilon}} & \overline{\overline{\alpha}}
 \\  \overline{\overline{\beta }} & \overline{\overline{\mu}} \end{array}
\right]\cdot \left[ \begin{array}{c} \mathbf{E} \\ \mathbf{H}\end{array} \right] 
\f
(see e.g., \cite{basic}). Obviously, only a limited range of material parameters is accessible in natural materials, and researchers have made significant efforts in synthesizing artificial composite materials with novel electromagnetic properties, not found in any natural substance. This is achieved by
engineering electrically small particles which respond to electromagnetic excitations as electric and magnetic dipoles in the desired way. The most general bianisotropic material can be conceptually realized as a mixture of many small inclusions modeled by the most general bianisotropic relations between the induced moments and exciting fields:
\e
\left[ \begin{array}{c} \mathbf{p} \\ \mathbf{m}\end{array} \right]
=\left[ \begin{array}{cc} \overline{\overline{\alpha}}_{\rm ee} & \overline{\overline{\alpha}}_{\rm em}
 \\  \overline{\overline{\alpha}}_{\rm me} & \overline{\overline{\alpha}}_{\rm mm} \end{array}
\right]\cdot \left[ \begin{array}{c} \mathbf{E} \\ \mathbf{H}\end{array} \right] .
\label{eq:a}\f
Here $\mathbf{p}$ and $\mathbf{m}$
are the induced electric and magnetic dipole moments, respectively, and $\overline{\overline{\alpha}}_{\rm ij}$ are the polarizability dyadics. 
In the modern literature this concept is called the \emph{metamaterial} concept, and small engineered inclusions in these composites are called \emph{meta-atoms}. In the past, significant results have been achieved in realizing artificial media with strong chirality (reciprocal magnetoelectric coupling measured by the trace of $\overline{\overline{\alpha}}_{\rm em}=-\overline{\overline{\alpha}}_{\rm me}^T$),
and in engineering permittivity and permeability  (values of $\overline{\overline{\alpha}}_{\rm ee}$ and $\overline{\overline{\alpha}}_{\rm mm}$), creating artificial magnetics and double-negative media. 

In this paper, we will focus on the problem of realization and optimization the magneto-electric coefficients which determine nonreciprocal mechanisms of magnetoelectric coupling. The main motivation of this research are recent discoveries of extreme properties of nonreciprocal bianisotropic particles  \cite{scattering, optimal}, such as extreme asymmetry in scattering response (invisible from one of the directions, optimal absorption from the opposite direction, etc). It has been understood that planar arrays of small nonreciprocal bianisotropic particles with special values of the coupling coefficients  
 can be used as transparent absorbing boundaries \cite{Tretyakov1} (which were introduced earlier for the termination of computational domains in finite-element methods \cite{Balanis}). Furthermore, such arrays  can be applied for novel implementations of perfect electromagnetic isolators \cite{absorber}, twist polarizers \cite{twist}, thin-sheet phase shifters, and other devices. Artificial moving medium with properly chosen values of the material parameters can transform electromagnetic fields in a very general manner \cite{field-transforming}. However, at this time all  these applications are only theoretical predictions. Scatterers with the desired parameter values do not exist in nature, and there are no known realizations of nonreciprocal bianisotropic meta-atoms. In this paper we develop an analytical model of two nonreciprocal meta-atoms, which realize nonreciprocal field coupling of both fundamental classes: Tellegen and ``moving particle'' coupling \cite{basic}. The results show  that desired novel effects can be achieved in particles made of conventional materials and having reasonable dimensions.

Nonreciprocity of media in applied DC magnetic fields or media with spontaneous magnetization is a common place in classical electrodynamics. This nonreciprocity obviously implies anisotropy of the medium and is related to the off-diagonal components of the  permittivity (magneto-optical media, magnetized plasmas) or permeability (gyrotropic media, e.g. ferrites) dyadics. On the other hand, nonreciprocal effects in bianisotropic media due to nonreciprocal nature of magnetoelectric interactions in the medium (see e.g. in \cite{basic}) are very rare and very weak. For example, the Tellegen coupling has been observed in some antiferromagnetic crystals at low frequencies \cite{CrO2}. 
It is clear that the metamaterial concept is the only possible route towards realization of nonreciprocal particles offering the theoretically expected  extreme performance.
 
In 1948 B.D.H. Tellegen suggested an idea of an electromagnetic gyrator, a general nonreciprocal four-pole \cite{Tellegen}. In electromagnetics, the gyrator element corresponds to a nonreciprocal particle with nonzero trace of $\overline{\overline{\alpha}}_{\rm em}=\overline{\overline{\alpha}}_{\rm me}^T$, and to composite media (Tellegen's medium) performed as a random mixture of such particles. Ari Sihvola conceptualized a microscopic particle with permanent electric and magnetic dipole moments parallel to one another and linked by a non-electromagnetic force \cite{chibi}. Coupling of electric and magnetic polarizations in such composite is of nonreciprocal nature. However, practical possibilities to create the Tellegen particle and medium remain problematic. In 1996, E. O. Kamenetskii \cite{kamenetskii} suggested magneto-static ferrite resonators shaped as tablets with partial metallization (shaped as a strip) on one side of the tablet as a conceptual nonreciprocal bianisotropic particle. Ferrite disks of small sizes compared to the wavelength operate as very compact resonators of magneto-static waves because of dramatic shortening of these waves compared to the electromagnetic wave in free space at the same frequency. In that work Kamenetskii predicted that a chain of such tablets should operate as a nonreciprocal bianisotropic waveguide \cite{kamenetskii}. He has shown that it should be possible to obtain controllable coupling coefficient of nonreciprocal tablets simply by tuning the bias magnetic field. In 1998 V. Dmitriev \cite{Dmitriev} generalized the results by Kamenetskii to a hypothetic medium which represented an anisotropic variant of the Tellegen medium and showed topologies which correspond to the other fundamental class of nonreciprocal bianisotropic coupling: the ``moving'' particle. In 1998, S. A. Tretyakov \cite{Tretyakov2} introduced two designs of anisotropic nonreciprocal scatterers which presumably possess more significant resonant nonreciprocity of their polarizabilities than the tablets of Kamenetskii.

In 2003 the group of Tretyakov  experimentally confirmed the existence of nonreciprocal magnetoelectric coupling \cite{Sanmartin} in one of the configurations introduced in \cite{Tretyakov2}, but in that work the particle polarizability value was not determined, and there is no model to predict and optimize the particle response.  In the present paper we make an important step towards realization of nonreciprocal bianisotropic meta-atoms with desired values of the polarizabilities, developing an analytical model which allows us to predict the polarizability values of particles with given topology and dimensions.

\section{Theory}
The geometry of the considered artificial nonreciprocal bianisotropic particles \cite{Tretyakov2,1998} is shown in Fig.~\ref{fig:10004} (Tellegen-omega) and Fig.~\ref{fig:10003} (moving-chiral).
The operation principle of both of them is that a local (external with respect to the particle) high-frequency electric field excites currents in the metal wires, and the magnetic field, created by the electric current of these wires induces magnetic moment in the ferrite sphere. This way a nonreciprocal magnetoelectric coupling is realized. Likewise, local high-frequency magnetic field applied to the ferrite sample causes its high-frequency magnetization which in turn induces electric current and electric dipole moment in the metal wires.

The general relation between the local fields and induced dipole moments is given by  \r{eq:a}. 
Tellegen-omega and moving-chiral particles, as the suggested nonreciprocal bianisotropic particles, are uniaxial. Therefore, we can write the polarizability dyadics of these two particles in the form
\e
\begin{split}
&\overline{\overline{\alpha}}_{\rm ee} =\aeeo \overline{\overline{I}}_{\rm{t}}+\aeer \overline{\overline{J}}_{\rm{t}},\\
&\overline{\overline{\alpha}}_{\rm mm} =\ammo \overline{\overline{I}}_{\rm{t}}+\ammr \overline{\overline{J}}_{\rm{t}},\\
&\overline{\overline{\alpha}}_{\rm em} =\aemo \overline{\overline{I}}_{\rm{t}}+\aemr \overline{\overline{J}}_{\rm{t}},\\
&\overline{\overline{\alpha}}_{\rm me} =\ameo \overline{\overline{I}}_{\rm{t}}+\amer \overline{\overline{J}}_{\rm{t}}.
\end{split}
\label{eq:b}
\f
Here, $\overline{\overline{I}}_{\rm{t}}=\overline{\overline{I}}-\mathbf{z}_0\mathbf{z}_0$
is the transverse unit dyadic, $\overline{\overline{I}}$ is the 3D unit dyadic, and
$\overline{\overline{J}}_{\rm{t}}=\mathbf{z}_0\times\overline{\overline{I}}_{\rm{t}}$ is the vector-product operator.

\subsection{Tellegen-Omega Particle}
\begin{figure}[h!]
    \centering
    \subfigure[]
    {
        \includegraphics[width=7cm]{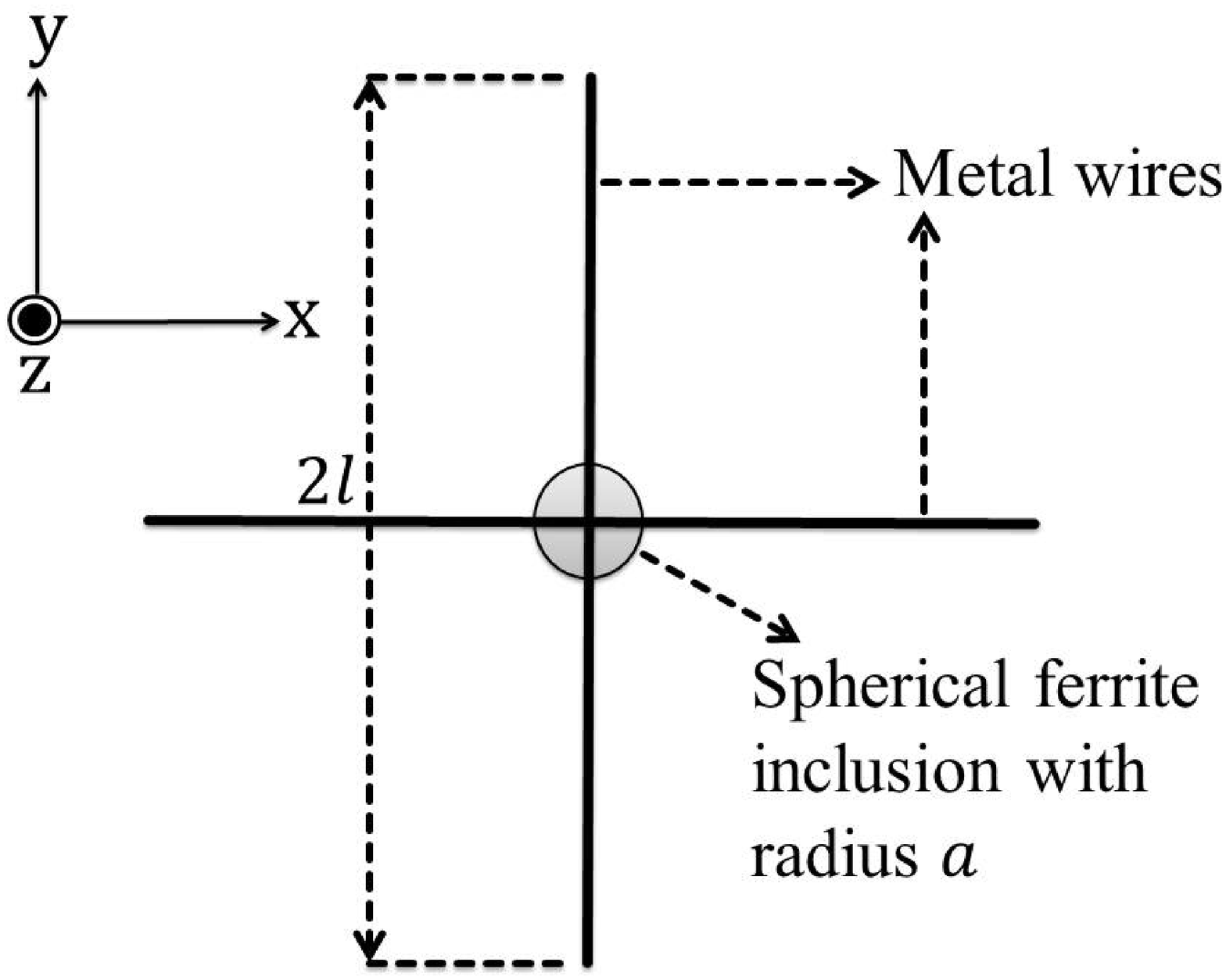}
        \label{fig:10004}
    }
    \subfigure[]
    {
        \includegraphics[width=7cm]{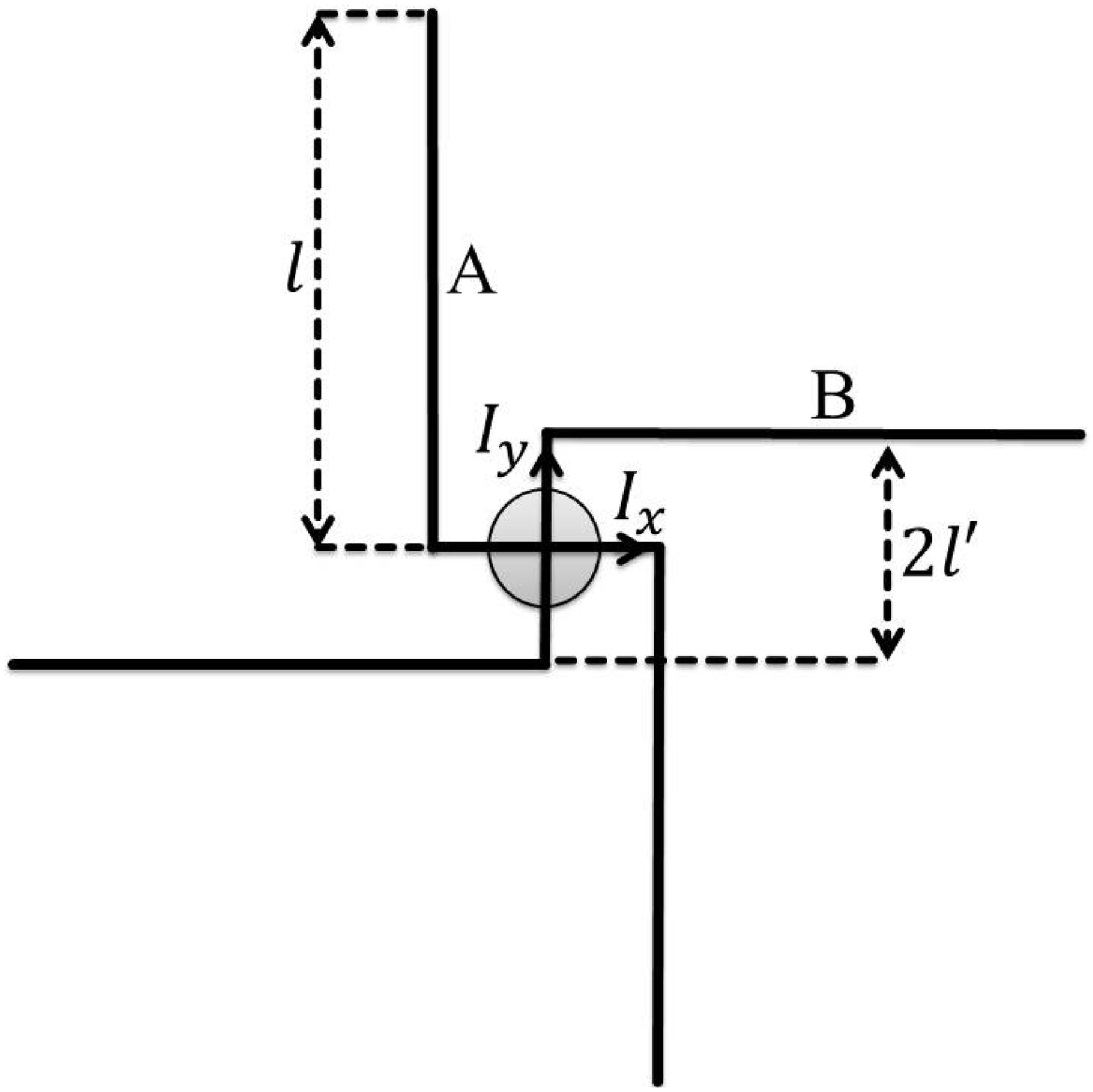}
        \label{fig:10003}
   }
        \caption{Geometry of the (a) Tellegen-omega particle and (b) moving-chiral particle. The bias magnetic field is directed along the $\rm{\mathbf{z}_0}$-axis.}
    \label{fig:jfd}
\end{figure}

Geometry of the particle is shown in Fig.~\ref{fig:10004}. 
Two metal wires (or strips) and a ferrite inclusion constitute the particle.


Tellegen-omega particle is one of those two particles which have been introduced as artificial nonreciprocal bianisotropic particles. It is clear from the particle name that it simultaneously exhibits two electro-magnetic coupling effects: nonreciprocal Tellegen and reciprocal omega coupling.

\subsubsection{Electric and magneto-electric polarizabilities}
\begin{figure}[h!]
  \begin{center}
    \includegraphics[width=9cm]{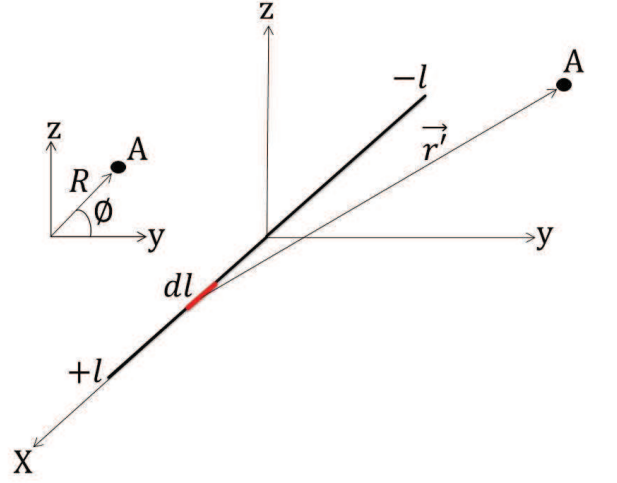}
    \caption{The wire along the $\mathbf{x}_0$-axis is excited by the incident electric field.}
    \label{fig:a}
  \end{center}
\end{figure}

We assume that the Tellegen-omega particle is excited by a uniform $\mathbf{x}_0$-directed electric field. This external field induces electric current in the metal wire which is in the direction of the electric field. Assuming that the wire length is much smaller than the wavelength $\left(l\ll \lambda \right)$, the current distribution in the wire is approximated as \cite{dipolemoment}
\e
I_x=I_{0x}\frac{\cos (kx)-\cos (kl)}{1-\cos (kl)}\approx I_{0x}\left(1-\frac{x^2}{l^2}\right),
\label{eq:c}
\f
where
\e
I_{0x}=\frac{2\tan \left(\displaystyle\frac{kl}{2}\right)}{kZ_{\rm{in}}}E\approx\frac{l}{Z_{\rm{in}}}E.
\label{eq:d}
\f
$E$ is the peak value of the incident electric field, $k$ is the free-space wavenumber, and $Z_{\rm{in}}$ represents the input impedance of a linear electric dipole antenna. For such an antenna, the input admittance can be expressed as\cite{harrison}
\e
\begin{array}{c}\displaystyle
Y_{\rm{in}}=2\pi j\frac{kl}{\eta\Psi}\left[1+k^2l^2\frac{F}{3}-jk^3l^3\frac{1}{3\left(\Omega-3\right)}\right],\vspace*{.3cm}\\\displaystyle
F=1+\frac{1.08}{\Omega-3},\hspace{.2cm}
\Omega=2\log\frac{2l}{r_0},\hspace{.2cm}
\Psi=2\log\frac{l}{r_0}-2,
\end{array}\label{eq:e}
\f
in which $\eta$ is the free-space wave impedance, $l$ is half of the length of the metal wire, and $r_0$ represents the wire radius. The induced electric current generates magnetic field. By applying the Biot-Savart law in the magneto-static approximation \cite{Jackson,Phillips,cheng}, the magnetic field close to the wire can be written as
\e
\begin{array}{l}\displaystyle
\mathbf{H}={\int_{-l}^{+l}\frac{\mathbf{I_{x}\,dl} \times \mathbf{r'}}{{r'}^3}}=\frac{I_{0x}\left(l^2-x^2\right)}{4\pi Rl^2}
\vspace*{.3cm}\\\displaystyle
\hspace*{1cm}\times
\left(\frac{l+x}{\sqrt{R^2+(l+x)^2}}+\frac{l-x}{\sqrt{R^2+(l-x)^2}}\right)
\vspace*{.3cm}\\\displaystyle
\hspace*{2cm}\times
\left(-\sin \phi\,\mathbf{y}_0+\cos \phi\,\mathbf{z}_0\right).
\end{array}\label{eq:f}
\f
As shown in Fig.~\ref{fig:a}, $\mathbf{r'}$ is the distance vector from a differential element to the observation point A. 
$R$ and $\phi$ are the cylindrical coordinates in the $\mathbf{y}_0\mathbf{z}_0$-plane.
The $\mathbf{y}_0$-component of the magnetic field, generated by the induced current in the wire, excites the ferrite sphere. As it is seen from (\ref{eq:f}), this component is not uniform over the ferrite sphere volume. Hence, it is necessary to take its volume average to find the equivalent uniform external magnetic field exciting the ferrite sphere, because the fundamental mode of magnetization oscillations is characterized by the uniform magnetization over the sphere volume. By assuming that the ferrite sphere is small, the averaged $\mathbf{y}_0$-component of the magnetic field can be achieved as
\e
\begin{array}{c}\displaystyle
H_{y-{\rm{average}}}=\frac{1}{V}{\int_{V} H_{y} \, dv}=\frac{2a^2}{3V}I_{0x}.
\end{array}\label{eq:h}
\f

For a ferrite sphere with the saturation magnetization along the $\mathbf{z}_0$-axis, the components of the magnetic moment are related to the external RF magnetic field as \cite{pozar}
\begin{equation}
\begin{array}{l}\displaystyle
m_x=V\left(\frac{\alpha}{\alpha^2+\beta^2}\chi_{xx}+\frac{\beta}{\alpha^2+\beta^2}\chi_{xy}\right)H_{xe}
\vspace*{.3cm}\\\displaystyle
\hspace*{1.5cm}
+V\left(\frac{\alpha}{\alpha^2+\beta^2}\chi_{xy}-\frac{\beta}{\alpha^2+\beta^2}\chi_{xx}\right)H_{ye},\vspace*{.3cm}\\\displaystyle
m_y=V\left(\frac{-\alpha}{\alpha^2+\beta^2}\chi_{xy}+\frac{\beta}{\alpha^2+\beta^2}\chi_{xx}\right)H_{xe}
\vspace*{.3cm}\\\displaystyle
\hspace*{1.5cm}
+V\left(\frac{\alpha}{\alpha^2+\beta^2}\chi_{xx}+\frac{\beta}{\alpha^2+\beta^2}\chi_{xy}\right)H_{ye}.
\end{array}\label{eq:i}
\end{equation}
Here
\e
\begin{array}{c}\displaystyle
\alpha=1+\frac{1}{3}\chi_{xx},\quad\quad\quad\beta=\frac{1}{3}\chi_{xy},
\end{array}\label{eq:j}
\f
and $H_{xe}$, $H_{ye}$ are the $\mathbf{x}_0$- and $\mathbf{y}_0$-components of the external magnetic field. Susceptibility elements are given by \cite{pozar, collin} for a lossy ferrite material. The averaged $\mathbf{y}_0$-component of the magnetic field, produced by the wire along the $\mathbf{x}_0$-axis, excites two non-zero components of the magnetic moment. These components are orthogonal to the bias field. The excited magnetic moment induces electric current in the wires. The $\mathbf{y}_0$- and $\mathbf{x}_0$-components of the magnetic moment cause the electric current induction on the $\mathbf{x}_0$- and $\mathbf{y}_0$-directed wires, respectively. The electric currents at the center of the wires, due to the magnetic moment, can be written as
\e
\begin{array}{l}\displaystyle
I_{0x}=\frac{1}{Z_{\rm{in}}}{\int_{-l}^{+l}E_{x}\left(1-\frac{\mid x \mid}{l}\right)dx}=+\frac{\xi}{Z_{\rm{in}}}m_y,\vspace*{.3cm}\\\displaystyle
I_{0y}=\frac{1}{Z_{\rm{in}}}{\int_{-l}^{+l}E_{y}\left(1-\frac{\mid y \mid}{l}\right)dy}=-\frac{\xi}{Z_{\rm{in}}}m_x,
\end{array}\label{eq:p}
\f
where $\xi$ is an unknown coefficient. After determining all the polarizabilities, the Onsager-Casimir principle \cite{onsager0, casimir, onsager} will allow us to determine this unknown coefficient. $E_x$ and $E_y$ represent the tangential to the wires electric field components generated by the magnetic moment. The excited wire along the $\mathbf{y}_0$-axis produces $\mathbf{x}_0$-directed magnetic field. Similarly to the adjacent wire, the volume average of this component of the magnetic field over the ferrite sphere volume can be expressed as
\e
\begin{array}{c}\displaystyle
H_{x-{\rm{average}}}=-\frac{2a^2}{3V}I_{0y}.
\end{array}\label{eq:q}
\f

Defining
\e
\begin{split}
&C_{xx}=C_{yy}\triangleq\frac{2a^{2}}{3}\left(\frac{\alpha}{\alpha^2+\beta^2}\chi_{xx}+\frac{\beta}{\alpha^2+\beta^2}\chi_{xy}\right),\\
&C_{xy}=-C_{yx}\triangleq\frac{2a^{2}}{3}\left(\frac{\alpha}{\alpha^2+\beta^2}\chi_{xy}-\frac{\beta}{\alpha^2+\beta^2}\chi_{xx}\right),
\end{split}
\label{eq:r}
\f
and considering \r{eq:h}, (\ref{eq:i}), and \r{eq:q}, the magnetic moment components in terms of the electric currents at the center of the wires can be written as
\e
\begin{array}{c}\displaystyle
m_x=C_{xy}I_{0x}-C_{xx}I_{0y},\quad\displaystyle
m_y=C_{yy}I_{0x}-C_{yx}I_{0y}.
\end{array}\label{eq:s}
\f

\begin{figure}[h!]
  \begin{center}
    \includegraphics[width=8cm]{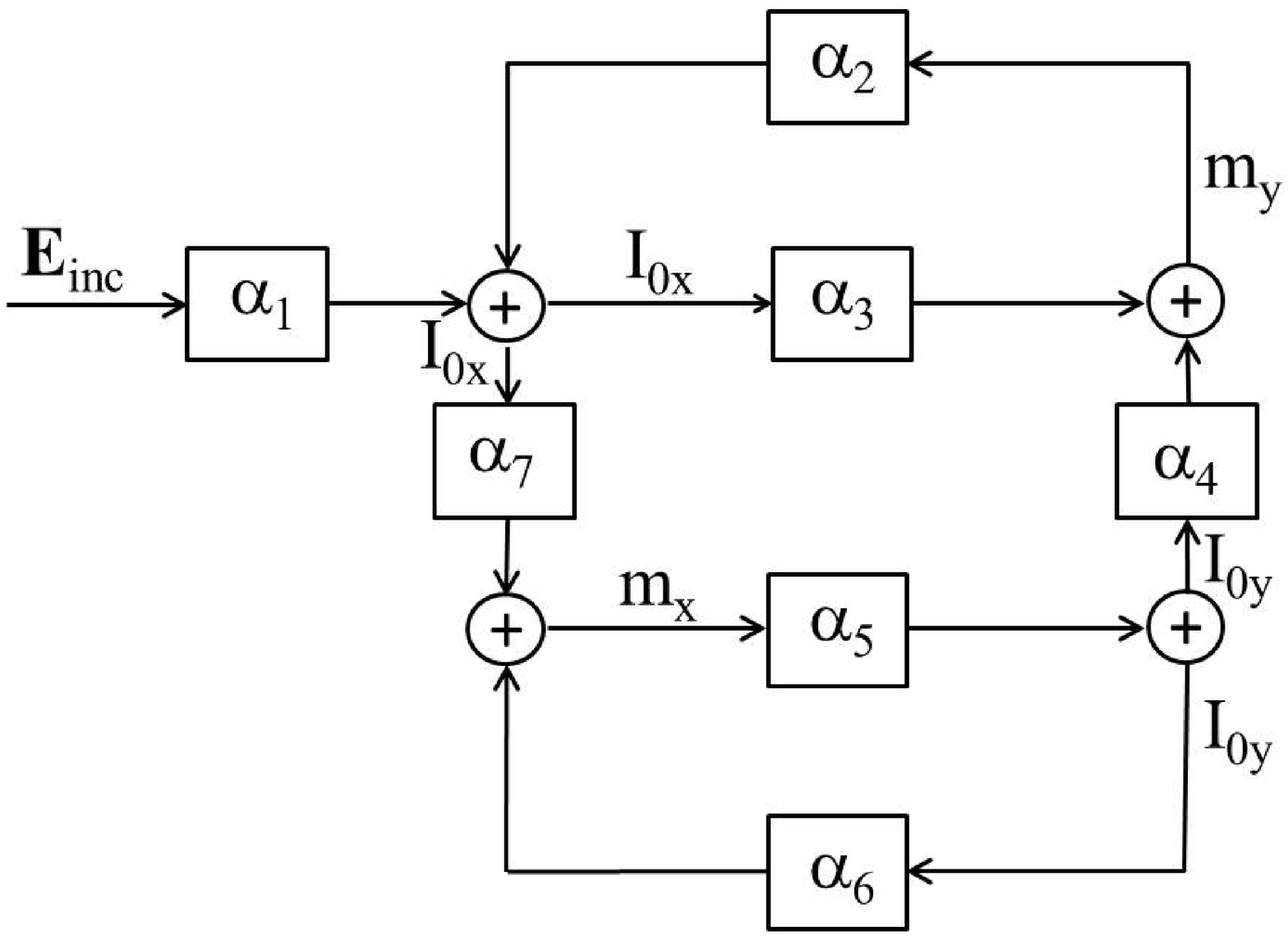}
    \caption{Block diagram of coupling between the ferrite inclusion and the metal wires in the Tellegen-omega particle in presence of an incident electric field in the $\mathbf{x}_0$-direction.}
    \label{fig:b}
  \end{center}
\end{figure}

As a result, it can be stated that there is a cyclic action, meaning that the electric currents excite the ferrite inclusion, and at the same time the magnetic moment excites the wires. This process can be modeled by a block diagram which can illustrate the relations between the currents and the magnetic moment components. Fig.~\ref{fig:b} shows the corresponding coupling block diagram.
Also, the cycle can be modeled by the following equations:
\e
\begin{split}
&I_{0x}=\alpha_{1}E+\alpha_{2}m_y,\hspace{.3cm}\displaystyle
m_y=\alpha_{3}I_{0x}+\alpha_{4}I_{0y},\\
&I_{0y}=\alpha_{5}m_{x},\hspace{1.4cm}\displaystyle
m_x=\alpha_{6}I_{0y}+\alpha_{7}I_{0x},
\end{split}
\label{eq:t}
\f
where
\e
\begin{array}{c}\displaystyle
\alpha_{1}\approx \frac{l}{Z_{\rm{in}}},\quad \alpha_{2}=-\alpha_{5}=\frac{\xi}{Z_{\rm{in}}},\quad \alpha_{3}=C_{yy},\vspace*{.3cm}\\ \alpha_{4}=-C_{yx},\quad \alpha_{6}=-C_{xx},\quad \alpha_{7}=C_{xy}.
\end{array}\label{eq:u}
\f
Knowing that the electric dipole moments and the electric currents at the center of the electrically small short-circuit wires are related to each other as\cite{dipolemoment}
\e
\begin{array}{c}\displaystyle
p_{x}\approx\frac{4l}{j3\omega}I_{0x},\quad\quad\quad
p_{y}\approx\frac{4l}{j3\omega}I_{0y},
\end{array}\label{eq:v}
\f
and solving \r{eq:t}, the electric and magneto-electric polarizabilities can be written as
\e
\begin{split}
&\aeeo=\frac{4l\alpha_1\left(1-\alpha_5\alpha_6\right)}{j3\omega(1-\alpha_2\alpha_3-\alpha_5\alpha_6+\alpha_2\alpha_3\alpha_5\alpha_6-\alpha_2\alpha_4\alpha_5\alpha_7)},\\
&\aeer=\frac{4l\alpha_1\alpha_5\alpha_7}{j3\omega(1-\alpha_2\alpha_3-\alpha_5\alpha_6+\alpha_2\alpha_3\alpha_5\alpha_6-\alpha_2\alpha_4\alpha_5\alpha_7)},\\
&\ameo=\frac{\alpha_1\alpha_7}{1-\alpha_2\alpha_3-\alpha_5\alpha_6+\alpha_2\alpha_3\alpha_5\alpha_6-\alpha_2\alpha_4\alpha_5\alpha_7},\\
&\amer=\frac{\alpha_1\alpha_3\left(1-\alpha_5\alpha_6\right)+\alpha_1\alpha_4\alpha_5\alpha_7}{1-\alpha_2\alpha_3-\alpha_5\alpha_6+\alpha_2\alpha_3\alpha_5\alpha_6-\alpha_2\alpha_4\alpha_5\alpha_7}.
\end{split}
\label{eq:w}
\f

It is important to know that the incident electric field can be considered as a uniform external field for the ferrite sphere as a homogeneous dielectric sphere which has the relative permittivity $\varepsilon_r$. Therefore, an electric dipole moment is induced parallel to the incident field. The absolute value of the moment is given (in the quasi-static approximation) by (e.g., \cite{sihvola})
\e
\begin{array}{c}\displaystyle
p=4\pi a^3\varepsilon_0\frac{\varepsilon_r-1}{\varepsilon_r+2}E,
\end{array}\label{eq:dielectric_sphere}
\f
where $\varepsilon_0$ is the permittivity of free space. Hence, there is an extra electric polarizability which should be added to the co-component of the electric polarizability in \r{eq:w}.

\subsubsection{Magnetic and electro-magnetic polarizabilities}
\begin{figure}[h!]
  \begin{center}
    \includegraphics[width=8cm]{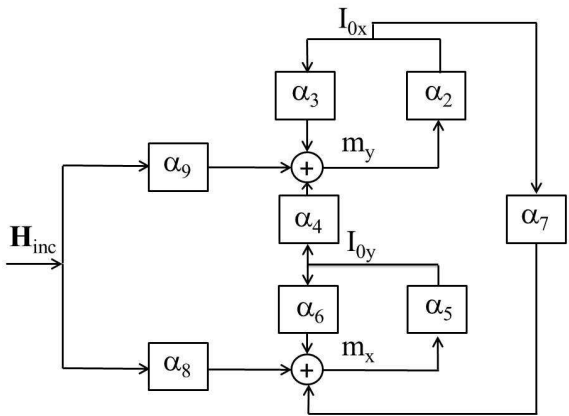}
    \caption{Block diagram of coupling between the ferrite inclusion and the metal wires in the Tellegen-omega particle in presence of an incident magnetic field in the $\mathbf{x}_0$-direction.}
    \label{fig:c}
  \end{center}
\end{figure}

To derive the magnetic and electro-magnetic polarizabilities, we assume that there is a high-frequency incident magnetic field in the plane of the particle, for example in the $\mathbf{x}_0$-direction. This field, with the peak value $H$, can excite the magnetic moment of the ferrite sphere. The excited magnetic moment induces electric current on the metal wires. Similarly to the previous process considered above, a coupling cycle is formed, because the induced electric currents excite the magnetic moment of the ferrite sphere. The following equations properly explain the cycle as
\e
\begin{array}{c}\displaystyle
I_{0x}=\alpha_{2}m_{y},\quad
m_y=\alpha_{3}I_{0x}+\alpha_{4}I_{0y}+\alpha_{9}H,\vspace*{.3cm}\\\displaystyle
I_{0y}=\alpha_{5}m_{x},\quad
m_x=\alpha_{6}I_{0y}+\alpha_{7}I_{0x}+\alpha_{8}H,
\end{array}\label{eq:x}
\f
where
\e
\begin{array}{c}\displaystyle
\alpha_{8}=2\pi aC_{xx},\quad\quad\quad\alpha_{9}=2\pi aC_{yx}.
\end{array}\label{eq:y}
\f
Solving \r{eq:x} and using \r{eq:v} give the magnetic and electro-magnetic polarizabilities as
\\
\e
\begin{split}
&\ammo=\frac{\alpha_8-\alpha_2\alpha_3\alpha_8+\alpha_2\alpha_7\alpha_9}{1-\alpha_2\alpha_3-\alpha_5\alpha_6+\alpha_2\alpha_3\alpha_5\alpha_6-\alpha_2\alpha_4\alpha_5\alpha_7},\\
&\ammr=\frac{\alpha_9}{1-\alpha_2\alpha_3-\alpha_5\alpha_6+\alpha_2\alpha_3\alpha_5\alpha_6-\alpha_2\alpha_4\alpha_5\alpha_7},\\
&\aemo=\frac{4l\alpha_2\alpha_9}{j3\omega(1-\alpha_2\alpha_3-\alpha_5\alpha_6+\alpha_2\alpha_3\alpha_5\alpha_6-\alpha_2\alpha_4\alpha_5\alpha_7)},\\
&\aemr=\frac{4l(\alpha_5\alpha_8\left(1-\alpha_2\alpha_3\right)+\alpha_2\alpha_5\alpha_7\alpha_9)}{j3\omega(1-\alpha_2\alpha_3-\alpha_5\alpha_6+\alpha_2\alpha_3\alpha_5\alpha_6-\alpha_2\alpha_4\alpha_5\alpha_7)}.
\end{split}
\label{eq:z}
\f
The corresponding coupling block diagram is shown in Fig.~\ref{fig:c}.

Now, by applying the Onsager-Casimir principle \cite{onsager}
\e
\overline{\overline{\alpha}}_{\rm{me}}\left(\rm{\mathbf{H}_0}\right)=-\overline{\overline{\alpha}}^T_{\rm{em}}\left(-\rm{\mathbf{H}_0}\right),
\label{eq:bb}
\f
it is possible to obtain the unknown coefficient $\xi$. $\mathbf{H}_0$ is the internal bias magnetic field and the superscript $T$ indicates the transpose operation. Using \r{eq:bb}, and considering \r{eq:w} and \r{eq:z}, after simple algebra the coefficient $\xi$ can be calculated as
\e
\xi=\frac{-j3\omega \mu_0}{8\pi a}.
\label{eq:aa}
\f

\subsection{Moving-Chiral Particle}
The other particle which has been introduced as the nonreciprocal bianisotropic particle is called moving-chiral. The geometry of the particle is illustrated in Fig.~\ref{fig:10003}. Similarly to the Tellegen-omega particle, the moving-chiral particle also consists of two metal wires which are placed on a ferrite inclusion which provides the necessary nonreciprocal response.

\subsubsection{Electric and magneto-electric polarizabilities}
An incident electric field in the $\mathbf{x}_0$-direction can excite both metal wires, because the shorter part of the wire A and the longer part of the wire B are parallel to the $\mathbf{x}_0$-axis. Assuming $l'\ll\lambda$, the small part of the wire A and the wire B have approximately uniform current distributions ($I_x$ and $I_y$, respectively).  The long parts of the wires are supposed to be still much smaller than the wavelength. Hence, the wires have approximately the following current distributions:
\begin{equation}
\begin{array}{c}\displaystyle
I_A = \left\{\begin{array}{rcl}
 I_{x}\left(1-\displaystyle\frac{y^2}{l^2}\right) & \mbox{for} & |y|>0 \\ I_x & \mbox{for}
& |x|<l',
\end{array}\right.
\vspace*{.3cm}\\\displaystyle
I_B = \left\{\begin{array}{rcl}
 I_y\left(1-\displaystyle\frac{x^2}{l^2}\right) & \mbox{for}
& |x|>0 \\ I_y & \mbox{for} & |y|<l',
\end{array}\right.
\end{array}\label{eq:M1}
\end{equation}
in which
\begin{equation}
\begin{array}{c}\displaystyle
I_{y}\approx \frac{l}{Z_{\rm{in}}}E,\qquad
I_{x}=\frac{2l'}{Z_{\rm{in}}}E.
\end{array}\label{eq:M2}
\end{equation}
The constant currents $I_{x}$ and $I_{y}$ become secondary sources which produce magnetic field for exciting the ferrite sphere. The $\mathbf{y}_0$-component of the magnetic field generated by $I_x$ and the $\mathbf{x}_0$-component of the magnetic field generated by $I_y$ have the most principal role in ferrite sphere excitation. Similarly to the theory of the Tellegen-omega particle, because of existing non-uniform external magnetic field within the ferrite sphere, calculation of the average of the field over the volume of the sphere should be done. Using the Biot-Savart law for the short part of wire A (where the current distribution is approximately uniform), and taking the average of the field over the volume of the ferrite sphere gives
\begin{equation}
\begin{array}{l}\displaystyle
H_{y-\rm{average}}=\frac{1}{V}{\int_{V} H_{y} \, dv}\vspace*{.3cm}\\\displaystyle
=\frac{I_x}{4\pi V}{\int_{\pi}^{2\pi}}{\int_0^{\pi}}{\int_0^{-2a\sin \theta \sin \phi} f(r,\,\theta , \,\phi)\,d r\, d\theta \, d\phi}=\frac{F}{V}I_x,\vspace*{.3cm}\\\displaystyle
f(r,\,\theta , \,\phi)=\left(\displaystyle\frac{l'+r\cos \theta}{\displaystyle\sqrt{r^2+l'^2+2rl'\cos \theta}}\right.
\vspace*{.3cm}\\\displaystyle 
\hspace*{2.5cm}
+ \left.\displaystyle\frac{l'-r\cos \theta}{\displaystyle\sqrt{r^2+l'^2-2rl'\cos \theta}}\right)\left(-r\sin \phi\right).
\end{array}\label{eq:M3}
\end{equation}
Calculation of the above integral is not straightforward. The value $F$ can be found numerically by, for instance, applying MATLAB simulator software. Similarly, the averaged $\mathbf{x}_0$-component of the external magnetic field due to the constant current $I_y$ can be expressed as
\begin{equation}
H_{x-{\rm{average}}}=-\frac{F}{V}I_y,
\label{eq:M4}
\end{equation}
in which the sign ''$-$'' implies that the produced magnetic field is opposite to the $\rm{\mathbf{x}_0}$-direction. By applying (\ref{eq:M3}), (\ref{eq:M4}), and defining the following coefficients
\begin{equation}
\begin{split}
&C_{xx}=C_{yy}\triangleq F\left(\frac{\alpha}{\alpha^2+\beta^2}\chi_{xx}+\frac{\beta}{\alpha^2+\beta^2}\chi_{xy}\right),\\
&C_{xy}=-C_{yx}\triangleq F\left(\frac{\alpha}{\alpha^2+\beta^2}\chi_{xy}-\frac{\beta}{\alpha^2+\beta^2}\chi_{xx}\right),
\end{split}
\label{eq:M6}
\end{equation}
Eq.~(\ref{eq:i}) reduces to
\begin{equation}
\begin{array}{c}\displaystyle
m_x=C_{xy}I_{x}-C_{xx}I_{y},\qquad
m_y=C_{yy}I_{x}-C_{yx}I_{y},
\end{array}\label{eq:M7}
\end{equation}which actually gives the magnetic moment in terms of the constant currents on the short parts of the wires A and B. It is important to appreciate that the excited magnetic moment induces electric current in the metal wires, because the $\mathbf{y}_0$-component of the magnetic moment produces an external $\mathbf{x}_0$-directed electric field which is tangential to the short part of the wire A, and therefore it can excite it. The same is true for the wire B due to the $\mathbf{x}_0$-component of the magnetic moment. The electric currents in the small parts of the the wires, due to the magnetic moment, can be written as
\begin{equation}
\begin{array}{c}\displaystyle
I_{x}=\frac{\xi}{Z_{\rm{in}}}m_y,\qquad
I_{y}=-\frac{\xi}{Z_{\rm{in}}}m_x,
\end{array}\label{eq:M8}
\end{equation}
where $\xi$ is an unknown coefficient. Similarly to what we did above for the Tellegen-omega particle, this coefficient will be found from the Onsager-Casimir principle.

\begin{figure}[h!]
  \begin{center}
    \includegraphics[width=8cm]{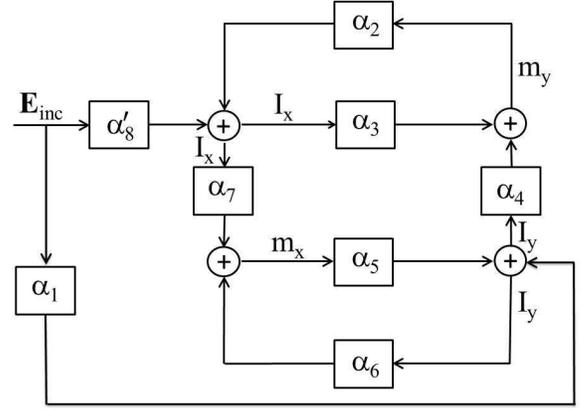}
    \caption{Block diagram of coupling between the metal wires and the ferrite inclusion in the moving-chiral particle in presence of an incident electric field in the $\mathbf{x}_0$-direction.}
    \label{fig:g}
  \end{center}
\end{figure}

Because the constant currents on the short parts of wires A and B excite the ferrite inclusion, and simultaneously the magnetic moment induces electric current on the wires, a coupling cycle is created, which is illustrated by the block diagram in Fig.~\ref{fig:g}.
The coupling cycle can be also expressed mathematically as the following relations:
\begin{equation}
\begin{array}{c}\displaystyle
I_x=\alpha'_{8}E+\alpha_{2}m_y,\qquad
m_y=\alpha_{3}I_{x}+\alpha_{4}I_y,\vspace*{.3cm}\\\displaystyle
I_y=\alpha_{5}m_{x}+\alpha_{1}E,\qquad
m_x=\alpha_{6}I_{y}+\alpha_{7}I_x,
\end{array}\label{eq:M9}
\end{equation}
where
\begin{equation}
\begin{array}{c}\displaystyle
\alpha_{1}\approx\frac{l}{Z_{\rm{in}}}, \quad \alpha_{2}=-\alpha_{5}=\frac{\xi}{Z_{\rm{in}}}, \quad \alpha_{3}=C_{yy},\vspace*{.3cm}\\\displaystyle
\alpha_{4}=-C_{yx}, \quad \alpha_{6}=-C_{xx}, \quad \alpha_{7}=C_{xy}, \quad \alpha'_{8}=\frac{2l'}{Z_{\rm{in}}}.
\end{array}\label{eq:M10}
\end{equation}
The electric dipole moments and the constant electric currents in the short parts of the wires are related to each other as
\begin{equation}
\begin{array}{c}\displaystyle
p_{x}\approx \frac{4l}{j3\omega}I_y+\frac{2l'}{j\omega}I_{x},\qquad
p_{y}\approx\frac{2l'}{j\omega}I_{y}-\frac{4l}{j3\omega}I_x.
\end{array}\label{eq:M11}
\end{equation}
Using (\ref{eq:M9}) and also considering (\ref{eq:M11}), the electric and magneto-electric polarizabilities can be written as
\begin{equation}
\begin{array}{c}\displaystyle
\alpha_{\rm{ee}}^{\rm{co}}=\frac{4l}{j3\omega}A_y+\frac{2l'}{j\omega}A_x,\qquad
\alpha_{\rm{ee}}^{\rm{cr}}=\frac{2l'}{j\omega}A_y-\frac{4l}{j3\omega}A_x,\vspace{.3cm}\\\displaystyle
\alpha_{\rm{me}}^{\rm{co}}=\frac{\alpha_1\alpha_6\left(1-\alpha_2\alpha_3\right)+\alpha_1\alpha_2\alpha_4\alpha_7+\alpha_7\alpha'_8}{1-\alpha_2\alpha_3-\alpha_5\alpha_6+\alpha_2\alpha_3\alpha_5\alpha_6-\alpha_2\alpha_4\alpha_5\alpha_7},\vspace*{.3cm}\\\displaystyle
\alpha_{\rm{me}}^{\rm{cr}}=\frac{\alpha_3\alpha'_8\left(1-\alpha_5\alpha_6\right)+\alpha_1\alpha_4+\alpha_4\alpha_5\alpha_7\alpha'_8}{1-\alpha_2\alpha_3-\alpha_5\alpha_6+\alpha_2\alpha_3\alpha_5\alpha_6-\alpha_2\alpha_4\alpha_5\alpha_7},
\end{array}\label{eq:M12}
\end{equation}
in which the coefficients $A_x$ and $A_y$ read
\begin{equation}
\begin{array}{c}\displaystyle
A_x=\frac{\alpha'_8\left(1-\alpha_5\alpha_6\right)+\alpha_1\alpha_2\alpha_4}{1-\alpha_2\alpha_3-\alpha_5\alpha_6+\alpha_2\alpha_3\alpha_5\alpha_6-\alpha_2\alpha_4\alpha_5\alpha_7},\vspace{.3cm}\\\displaystyle
A_y=\frac{\alpha_1\left(1-\alpha_2\alpha_3\right)+\alpha_5\alpha_7\alpha'_8}{1-\alpha_2\alpha_3-\alpha_5\alpha_6+\alpha_2\alpha_3\alpha_5\alpha_6-\alpha_2\alpha_4\alpha_5\alpha_7}.
\end{array}\label{eq:MOVING1}
\end{equation}

\subsubsection{Magnetic and electro-magnetic polarizabilities}
Most of the formulas given above can be used, and we only need to rewrite the relations between the constant currents and the magnetic moment as
\begin{equation}
\begin{array}{c}\displaystyle
I_x=\alpha_{2}m_{y},\qquad
m_y=\alpha_{3}I_{x}+\alpha_{4}I_y+\alpha_{9}H,\vspace{.3cm}\\\displaystyle
I_y=\alpha_{5}m_{x},\qquad
m_x=\alpha_{6}I_{y}+\alpha_{7}I_x+\alpha_{8}H.
\end{array}\label{eq:M13}
\end{equation}
\\
Here $H$ is the peak value of the high-frequency incident magnetic field. If we assume that this incident magnetic field has only an $\mathbf{x}_0$-component, then
\begin{equation}
\begin{array}{c}\displaystyle
\alpha_{8}=\frac{4\pi a^3}{3F}C_{xx}, \qquad \alpha_{9}=\frac{4\pi a^3}{3F}C_{yx}.
\end{array}\label{eq:M14}
\end{equation}

By applying (\ref{eq:M11}) and (\ref{eq:M13}), the magnetic and electro-magnetic polarizabilities can be expressed as
\begin{equation}
\begin{array}{c}\displaystyle
\alpha_{\rm{mm}}^{\rm{co}}=\frac{\alpha_8\left(1-\alpha_2\alpha_3\right)+\alpha_2\alpha_7\alpha_9}{1-\alpha_2\alpha_3-\alpha_5\alpha_6+\alpha_2\alpha_3\alpha_5\alpha_6-\alpha_2\alpha_4\alpha_5\alpha_7},\vspace{.3cm}\\\displaystyle
\alpha_{\rm{mm}}^{\rm{cr}}=\frac{\alpha_9}{1-\alpha_2\alpha_3-\alpha_5\alpha_6+\alpha_2\alpha_3\alpha_5\alpha_6-\alpha_2\alpha_4\alpha_5\alpha_7},\vspace{.3cm}\\\displaystyle
\alpha_{\rm{em}}^{\rm{co}}=\frac{4l}{j3\omega}B_y+\frac{2l'}{j\omega}B_x,\qquad
\alpha_{\rm{em}}^{\rm{cr}}=\frac{2l'}{j\omega}B_y-\frac{4l}{j3\omega}B_x,
\end{array}\label{eq:M15}
\end{equation}
where the coefficients $B_x$ and $B_y$ are given by
\begin{equation}
\begin{array}{c}\displaystyle
\hspace{-0.15cm}B_x=\frac{\alpha_2\alpha_9}{1-\alpha_2\alpha_3-\alpha_5\alpha_6+\alpha_2\alpha_3\alpha_5\alpha_6-\alpha_2\alpha_4\alpha_5\alpha_7},\vspace{.3cm}\\\displaystyle
\hspace{-0.15cm}B_y=\frac{\alpha_5\alpha_8\left(1-\alpha_2\alpha_3\right)+\alpha_2\alpha_5\alpha_7\alpha_9}{1-\alpha_2\alpha_3-\alpha_5\alpha_6+\alpha_2\alpha_3\alpha_5\alpha_6-\alpha_2\alpha_4\alpha_5\alpha_7}.
\end{array}\label{eq:MOVING2}
\end{equation}

As it is seen from (\ref{eq:M13}), the coupling block diagram for this case is completely similar to what is shown in Fig.~\ref{fig:c} for the Tellegen-omega particle. The coefficients $\alpha_2$ and $\alpha_5$, which show the effect of the magnetic moment on the metal wires, are determined by the Onsager-Casimir principle. After some algebraic manipulations we can get
\begin{equation}
\alpha_2=-\alpha_5=\mu_0\frac{\alpha_1\alpha_4+\alpha_3\alpha'_8}{\left(\displaystyle\frac{4l}{j3\omega}\right)\alpha_9-\left(\displaystyle\frac{2l'}{j\omega}\right)\alpha_8}.
\label{eq:M17}
\end{equation}

\section{Results and discussion}

Here we consider the polarizablities of two example particles, compare the analytical predictions with numerical simulations, and discuss the results.

\subsection{Tellegen-Omega Particle}
\begin{figure}[h!]
    \centering
    \subfigure[$\eta\aeeo$]
    {
        \includegraphics[width=4cm]{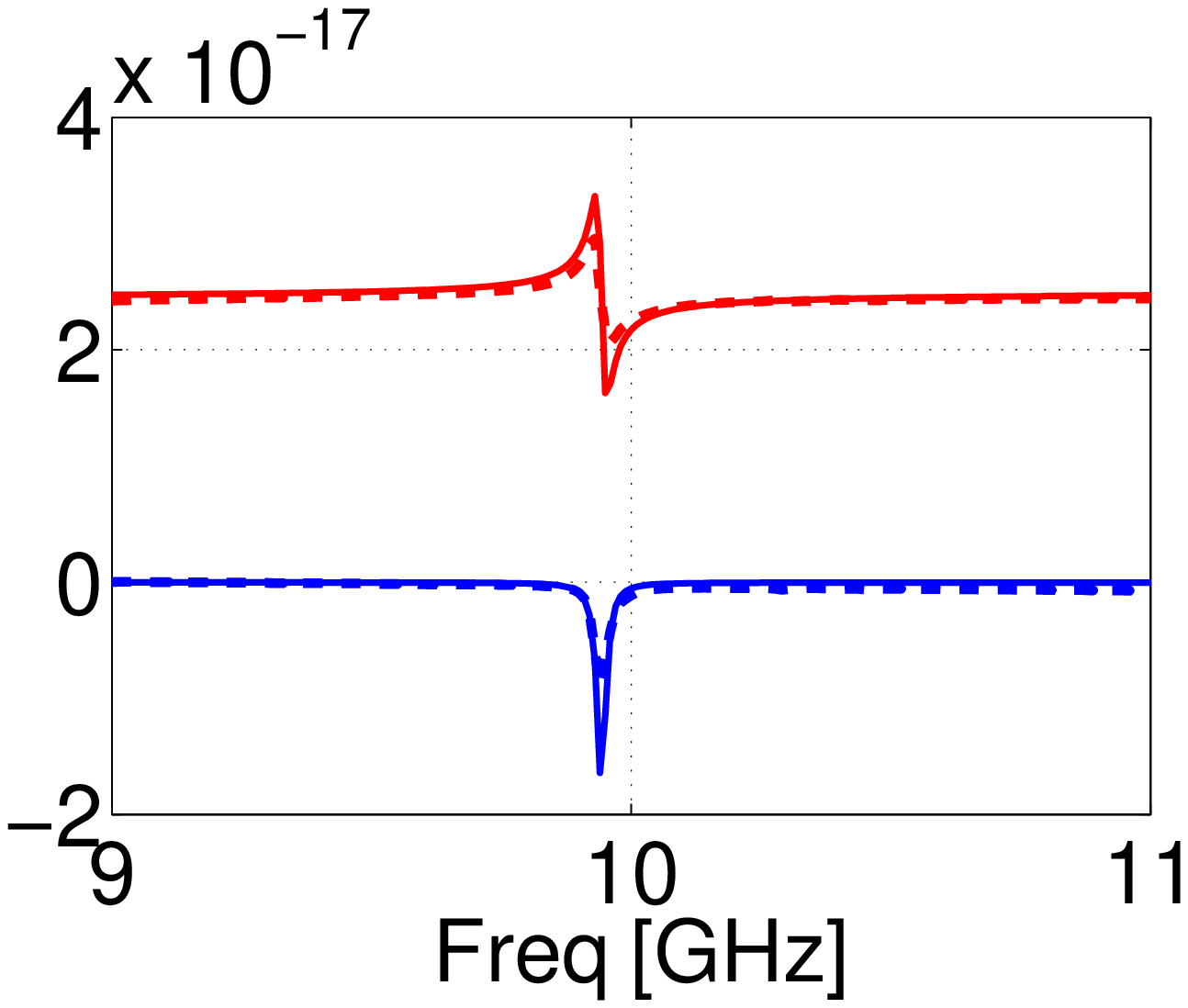}
        \label{fig:e_1}
    }
    \subfigure[$\eta\aeer$]
    {
        \includegraphics[width=4cm]{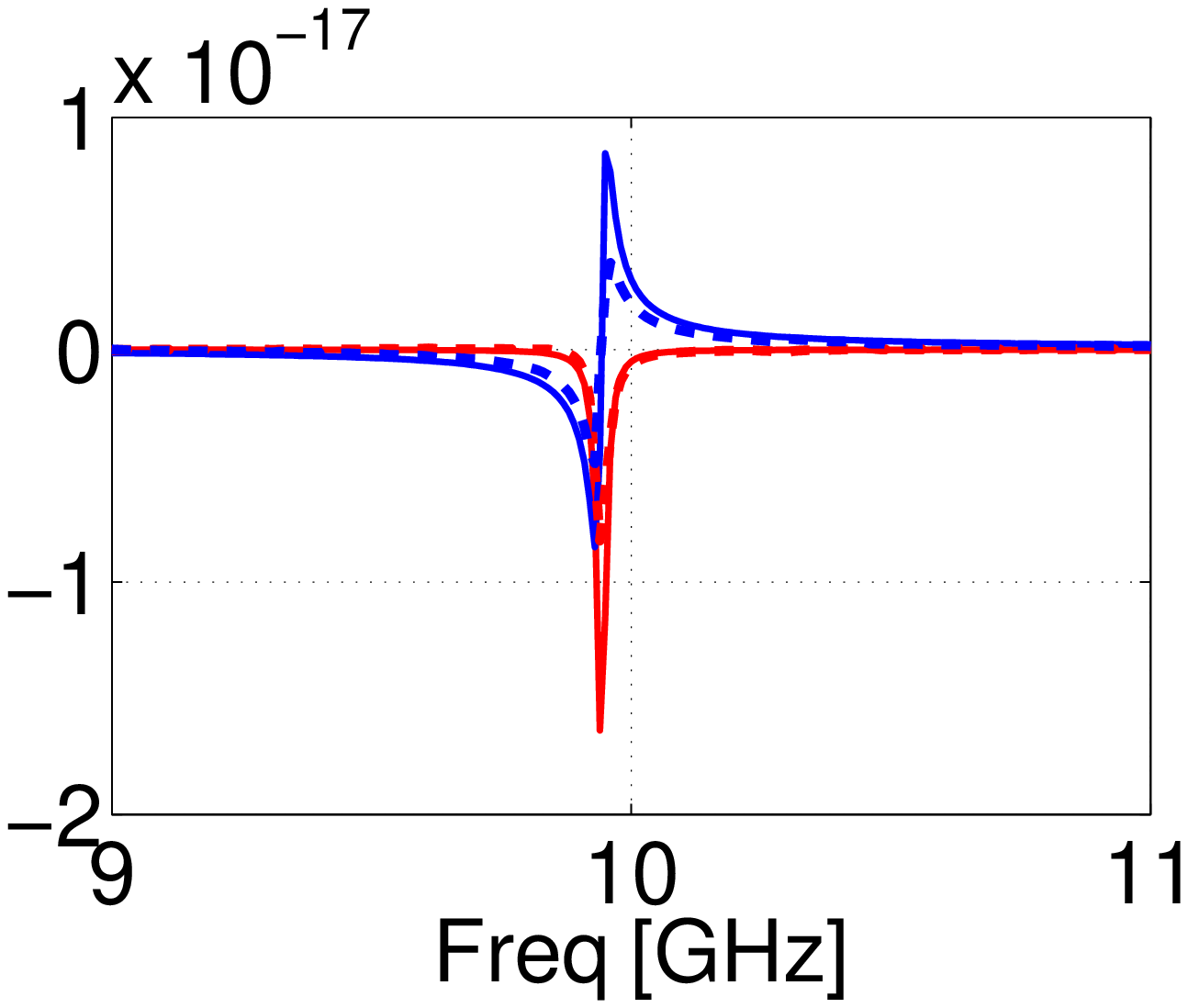}
        \label{fig:e_2}
   }
     \\
    \subfigure[$\aemo=\ameo$]
    {
        \includegraphics[width=4cm]{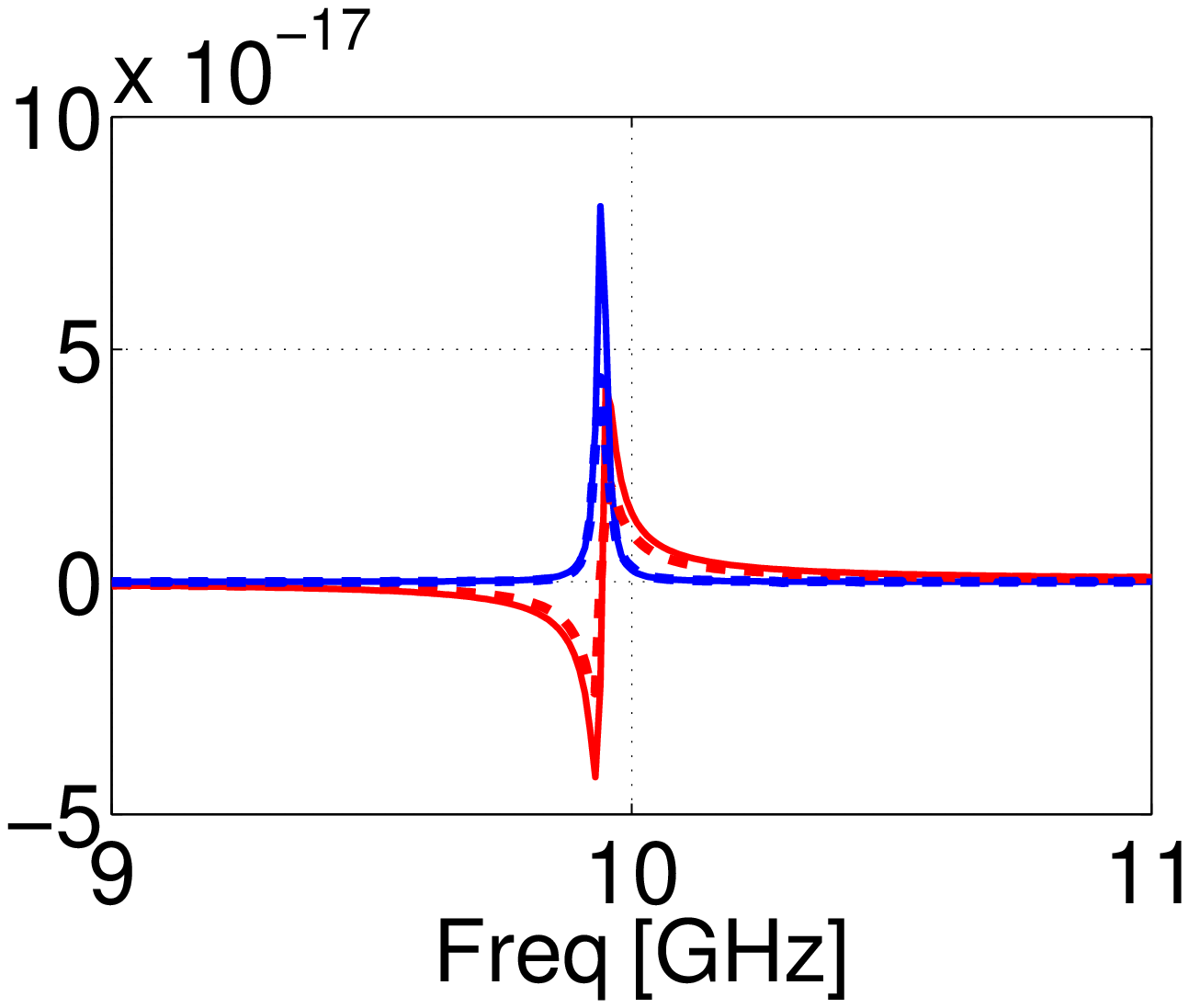}
        \label{fig:e_3}
    }
    \subfigure[$\aemr=\amer$]
    {
        \includegraphics[width=4cm]{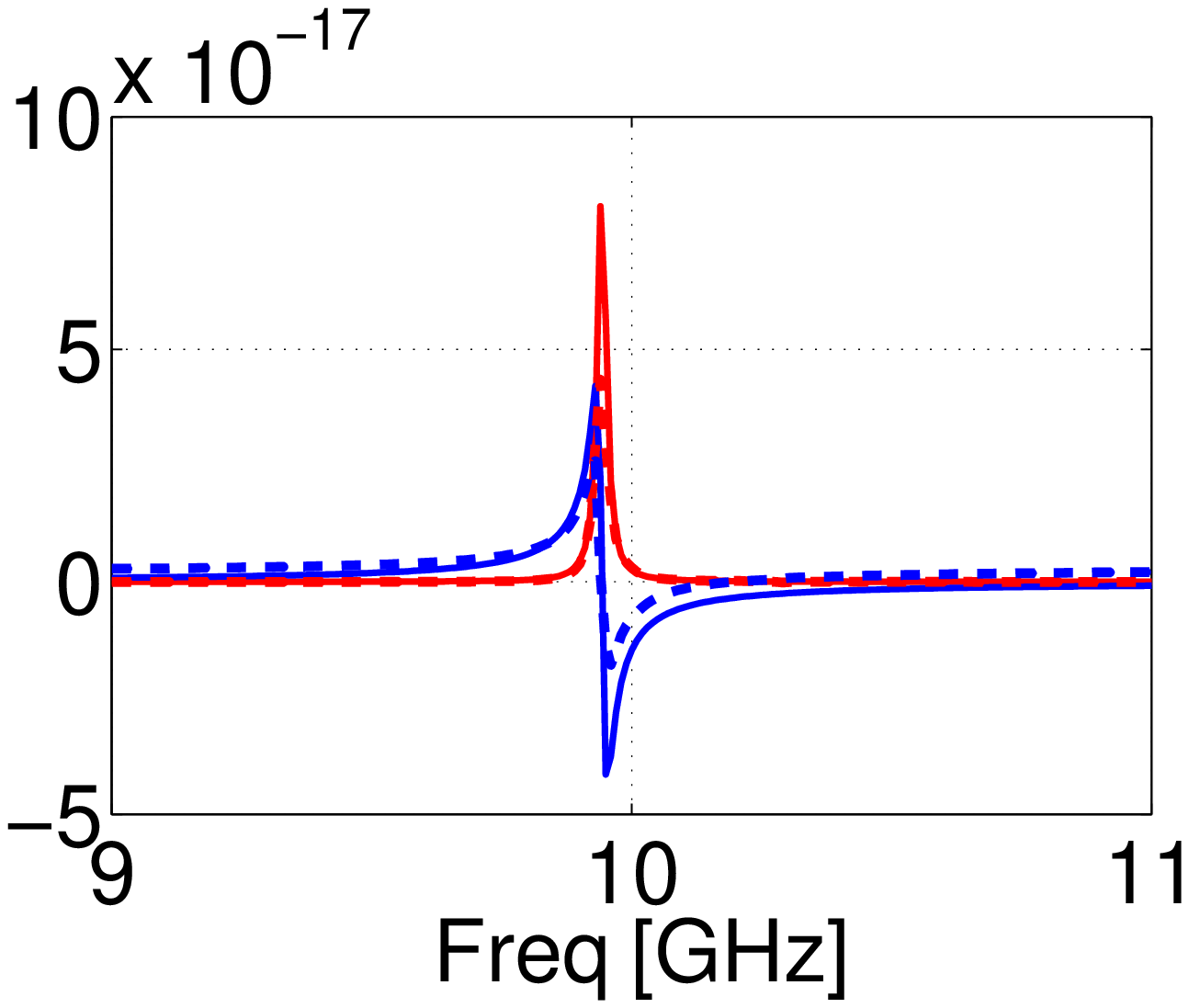}
        \label{fig:e_4}
    }
     \\
    \subfigure[$\displaystyle\frac{1}{\eta}\ammo$]
    {
        \includegraphics[width=4cm]{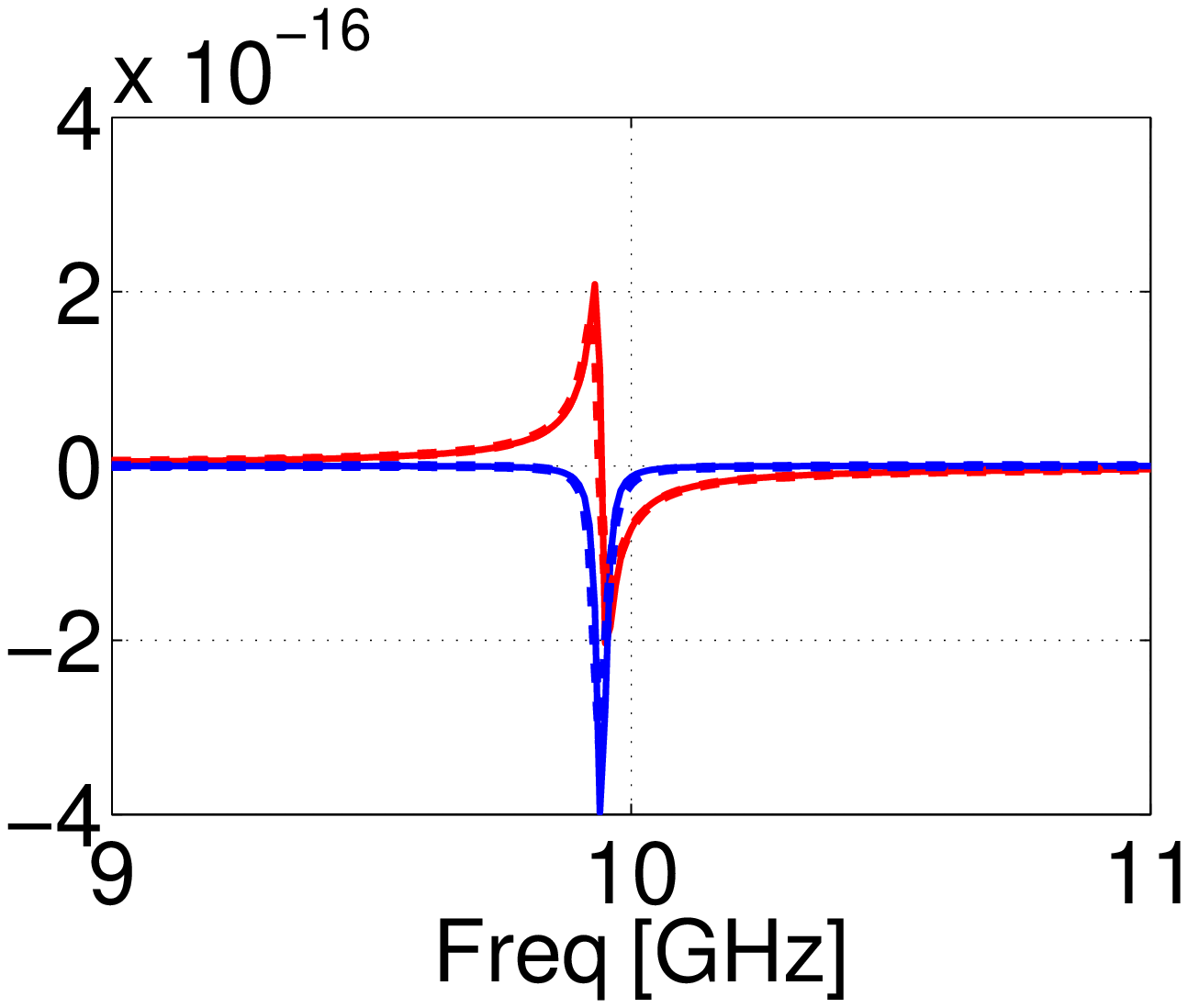}
        \label{fig:e_5}
    }
    \subfigure[$\displaystyle\frac{1}{\eta}\ammr$]
    {
        \includegraphics[width=4cm]{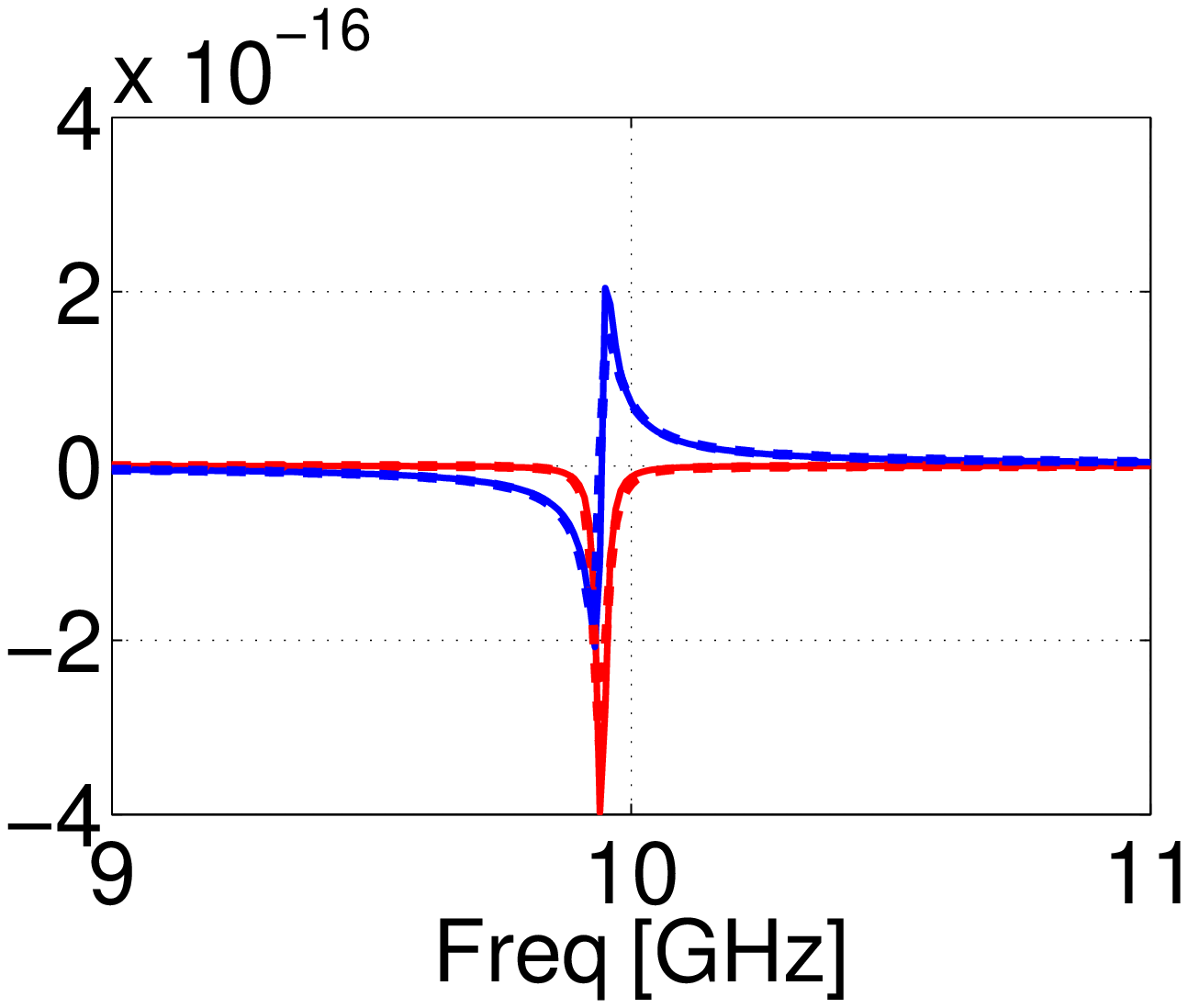}
        \label{fig:e_6}
    }
    \caption{Comparison of simulated and analytical polarizabilities of Tellegen-omega particle. Solid lines are the analytical polarizabilities and dashed lines are the simulated polarizabilities. Red and blue colors represent the real and imaginary parts of the polarizability, respectively.}
    \label{fig:e}
\end{figure}

As an example, we study the properties of the particles whose constituent material properties and dimensions are illustrated in Table~\ref{table:a}. 
\begin{table}[!htb]
\center
\begin{tabular}{|p{1.1cm}|p{1.4cm}|p{1.8cm}|p{1.3cm}|p{1.05cm}|p{1.05cm}|}
\hline
\textbf{Ferrite Material} & \textbf{Relative permittivity} & \textbf{Saturation Magnetization} & \textbf{Applied Bias Field} & \textbf{Sphere Radius } & \textbf{Wires Length}\\
\hline
YIG  & 15 & 1780 (gauss)   & 3570 ($\rm{Oe}$)   & 0.5 ($\rm{mm}$)  & 3.0 ($\rm{mm}$)\\
\hline
\end{tabular}
\caption{Tellegen-omega particle, properties and dimensions.}
\label{table:a}
\end{table}
For having the resonance frequency at 10 GHz, the bias magnetic field should be 3570 $\rm{Oe}$. The ferrite material is assumed to be moderately lossy with the damping factor equal to 0.001. The length of each wire is 3 mm which is one tenth of the wavelength. Hence, the current distribution considered in (\ref{eq:c}) would be a valid model. To confirm the analytical results, we have found the polarizabilities also numerically based on the approach developed in \cite{victorAsad}. Simulations have been done by the Finite Element Method (FEM), applying ANSYS HFSS software. Simulations can be also done using the Method of Moment (MOM). Fig.~\ref{fig:e} compares the analytical polarizabilities with the numerical ones. 
As it is seen, the resonance frequency is approximately 10 GHz, and the simulated and analytical results are fairly well matched.

\subsection{Moving-Chiral Particle}
\begin{figure}[h!]
    \centering
    \subfigure[$\eta\aeeo$]
    {
        \includegraphics[width=4cm]{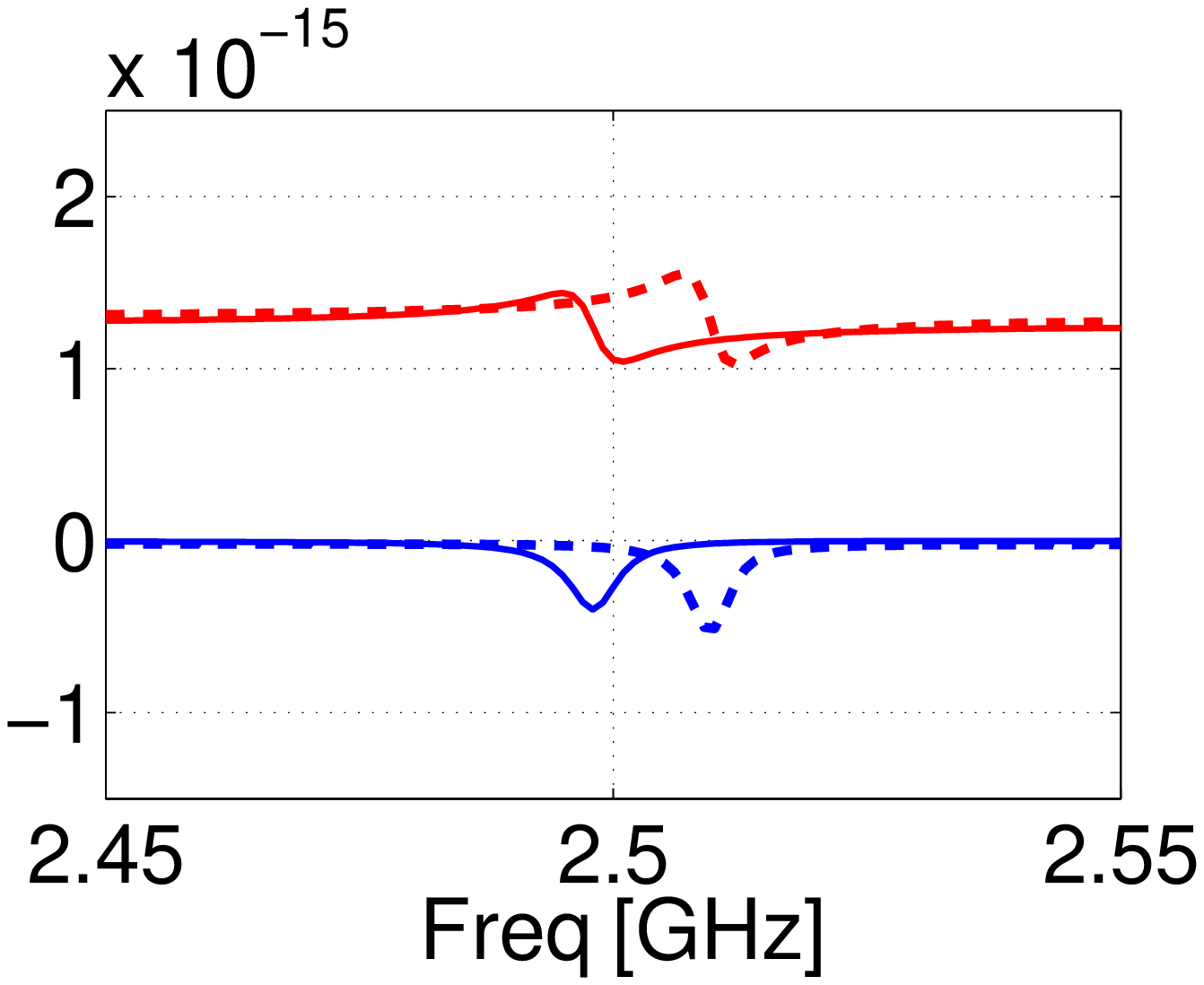}
        \label{fig:i_1}
    }
    \subfigure[$\eta\aeer$]
    {
        \includegraphics[width=4cm]{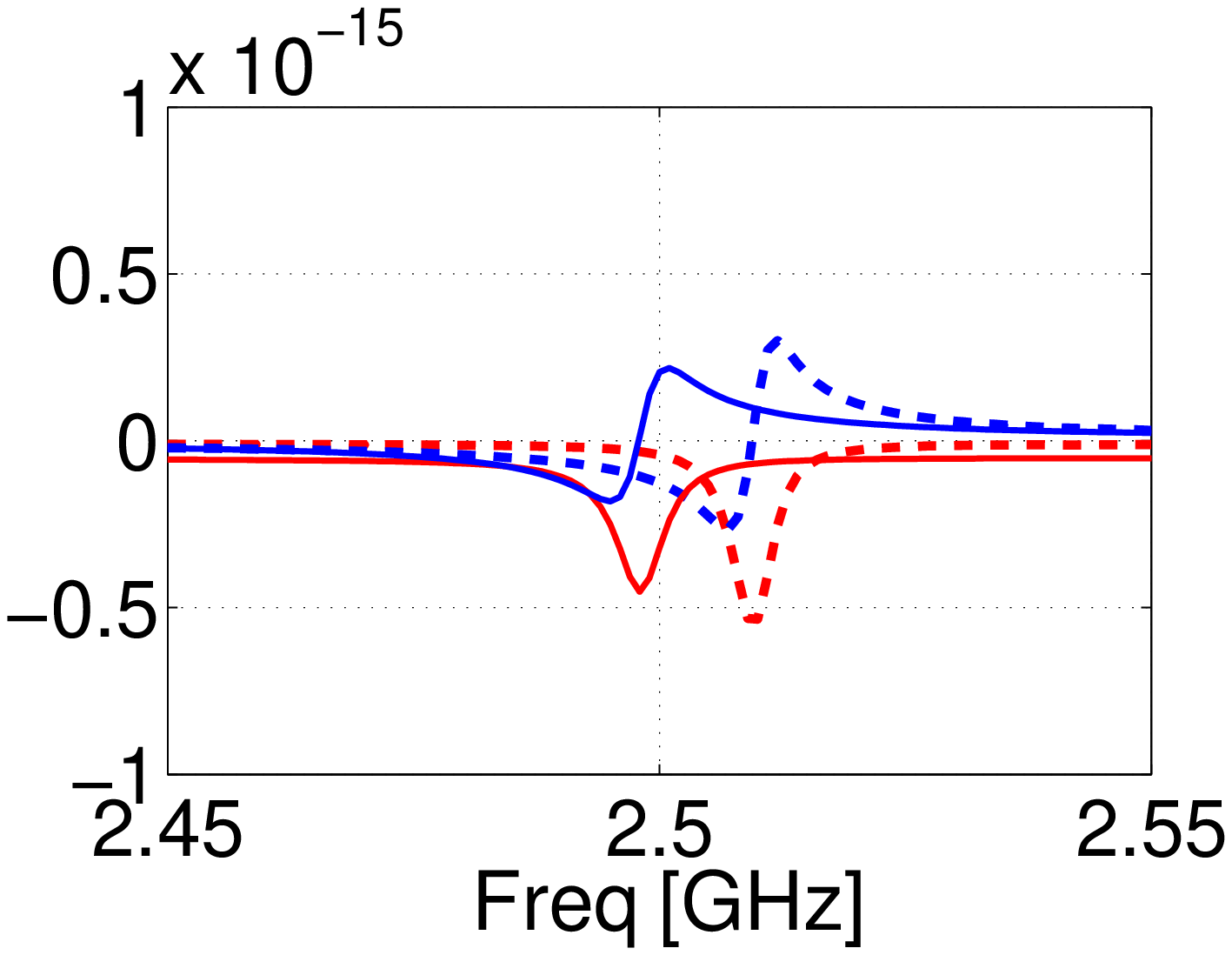}
        \label{fig:i_2}
   }
     \\
    \subfigure[$\displaystyle\frac{1}{\eta}\ammo$]
    {
        \includegraphics[width=4cm]{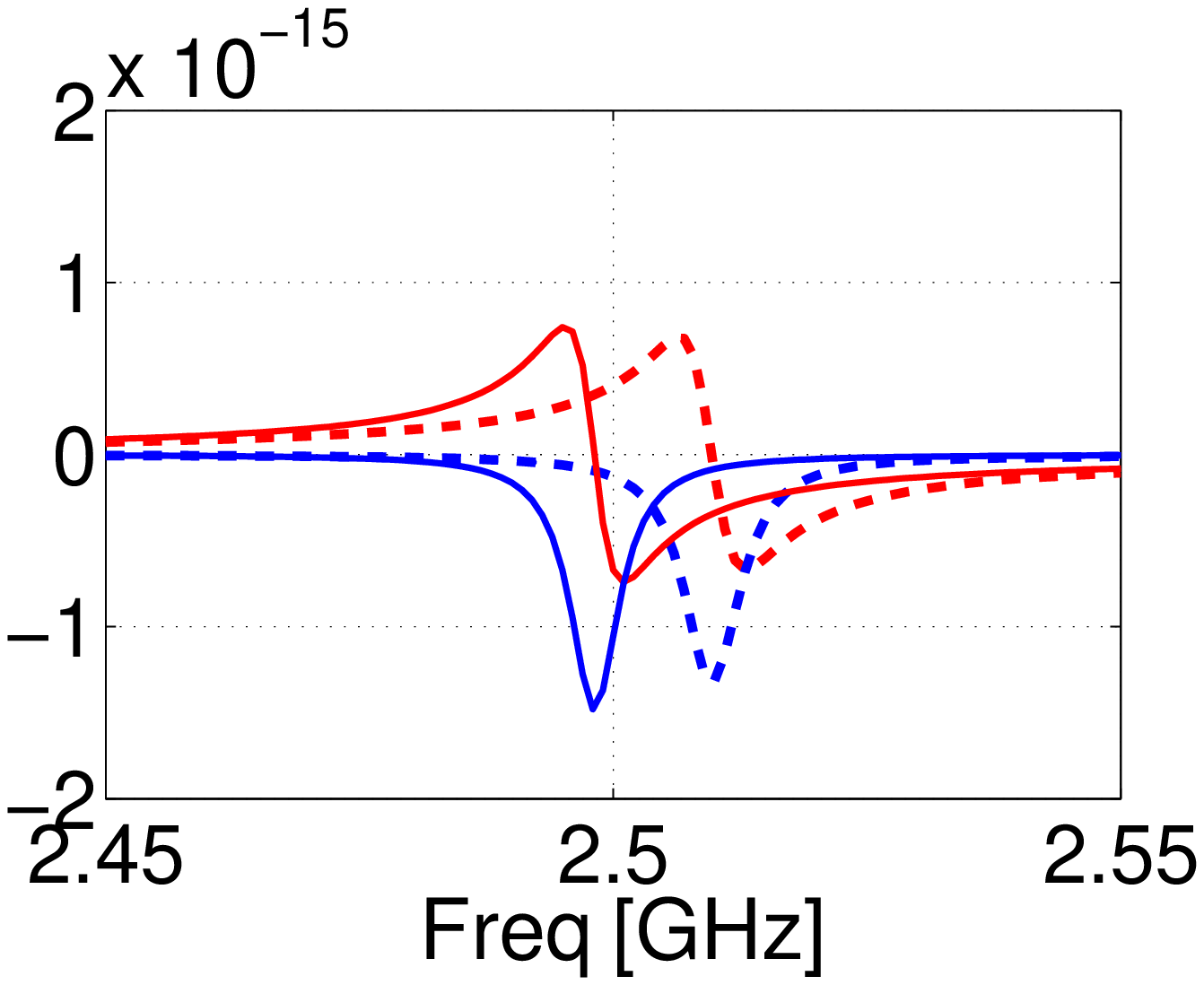}
        \label{fig:i_3}
    }
    \subfigure[$\displaystyle\frac{1}{\eta}\ammr$]
    {
        \includegraphics[width=4cm]{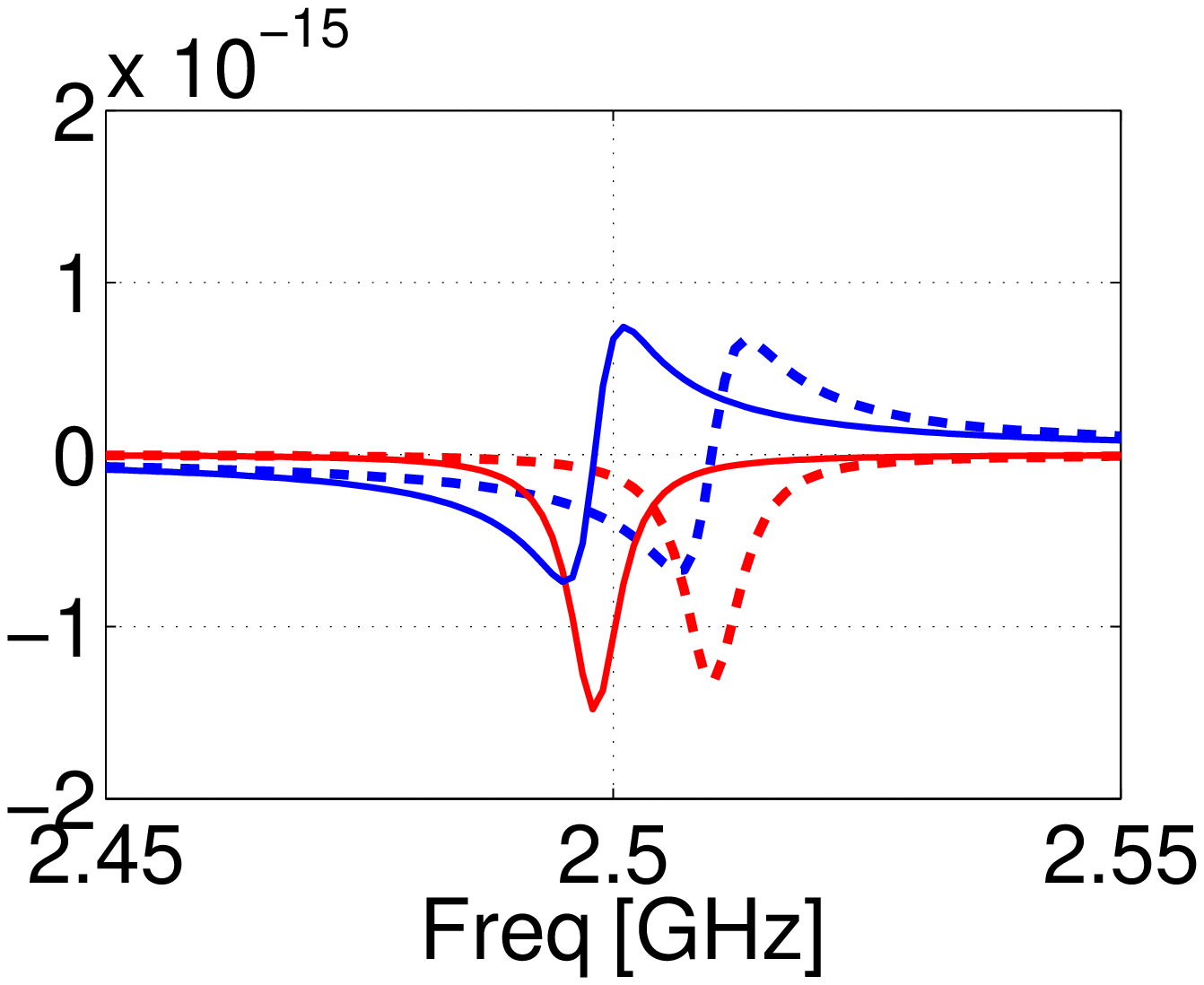}
        \label{fig:i_4}
    }
    \caption{Comparison of analytical and simulated (electric and magnetic) polarizabilities of the moving-chiral particle with $l=6.9\,\rm{mm}$ and $l'=0.6\,\rm{mm}$. Solid lines are the analytical polarizabilities and dashed lines are the simulated polarizabilities. Red and blue colors represent the real and imaginary parts of the polarizability, respectively.}
    \label{fig:i}
\end{figure}

\begin{figure}[h!]
    \centering
    \subfigure[$\aemo$]
    {
        \includegraphics[width=4cm]{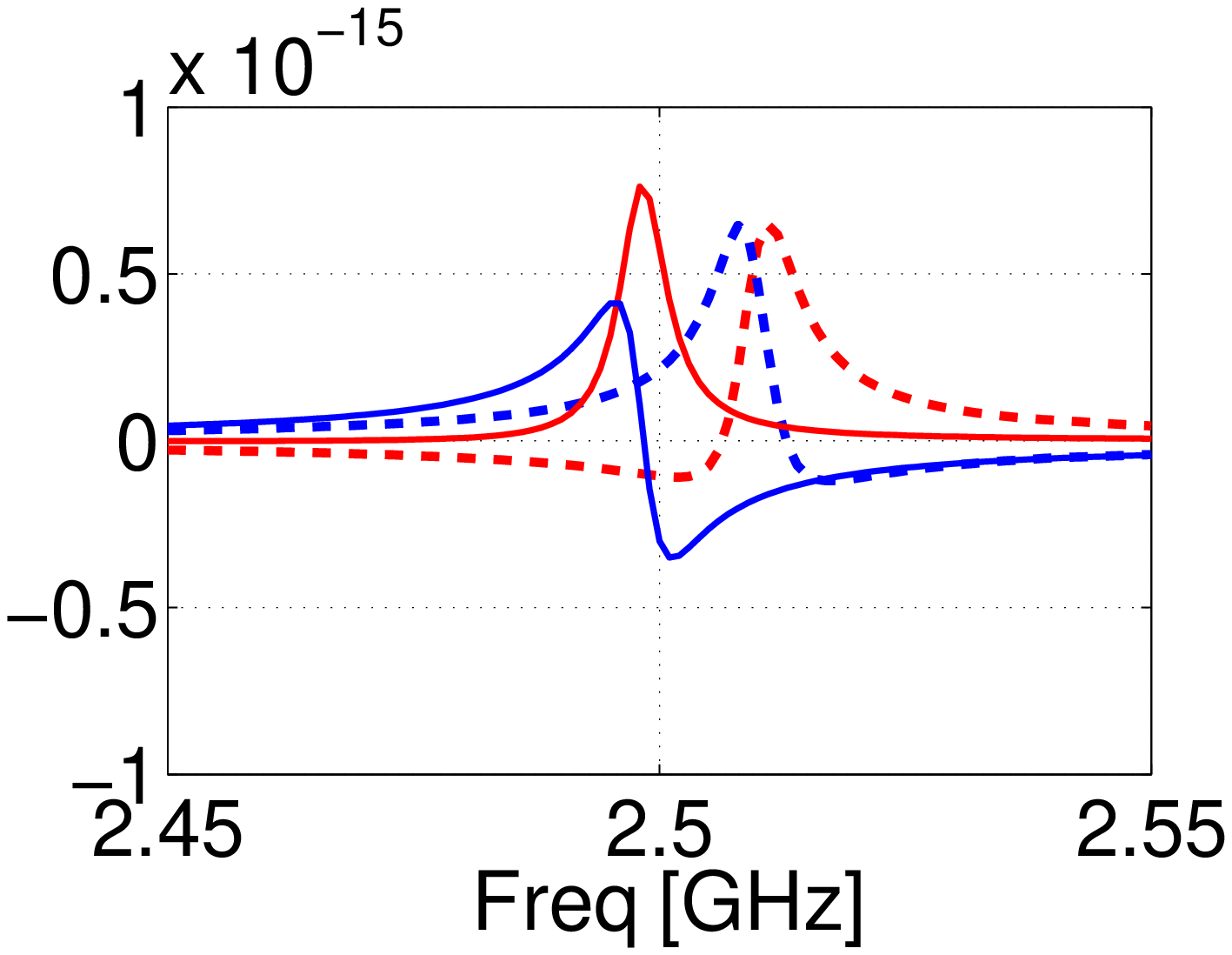}
        \label{fig:j_1}
    }
    \subfigure[$\ameo$]
    {
        \includegraphics[width=4cm]{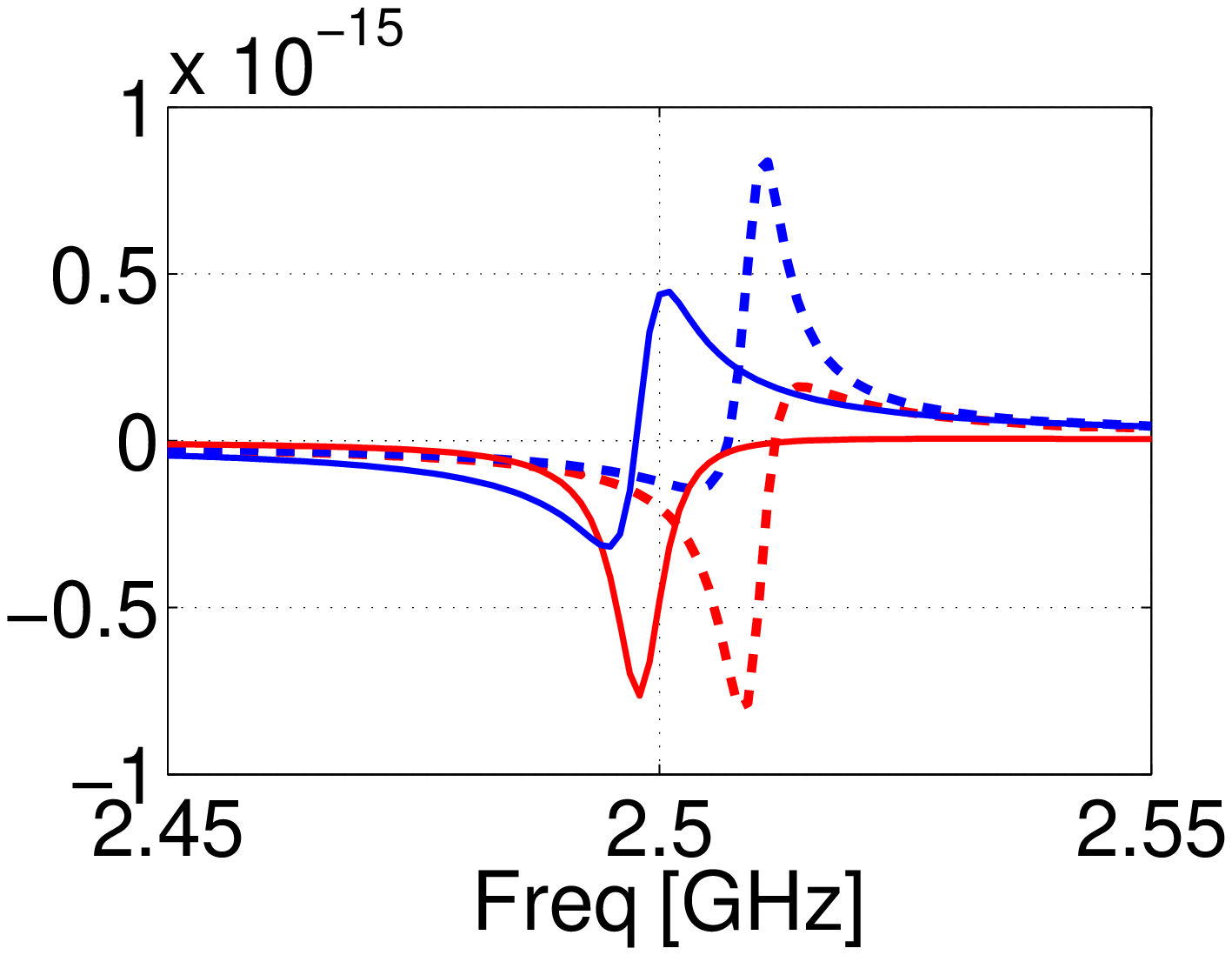}
        \label{fig:j_2}
   }
    \subfigure[$\aemr$]
    {
        \includegraphics[width=4cm]{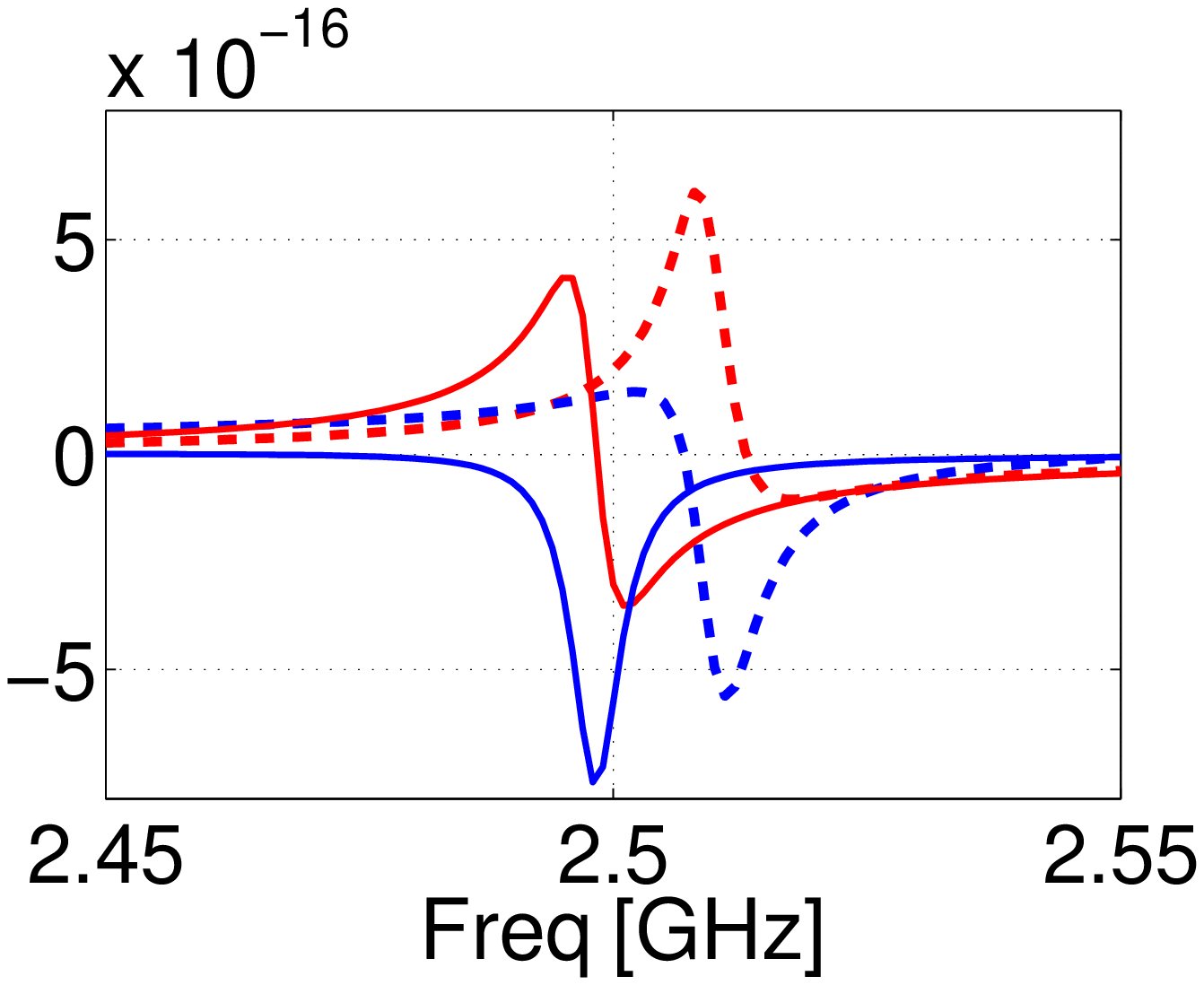}
        \label{fig:j_3}
    }
    \subfigure[$\amer$]
    {
        \includegraphics[width=4cm]{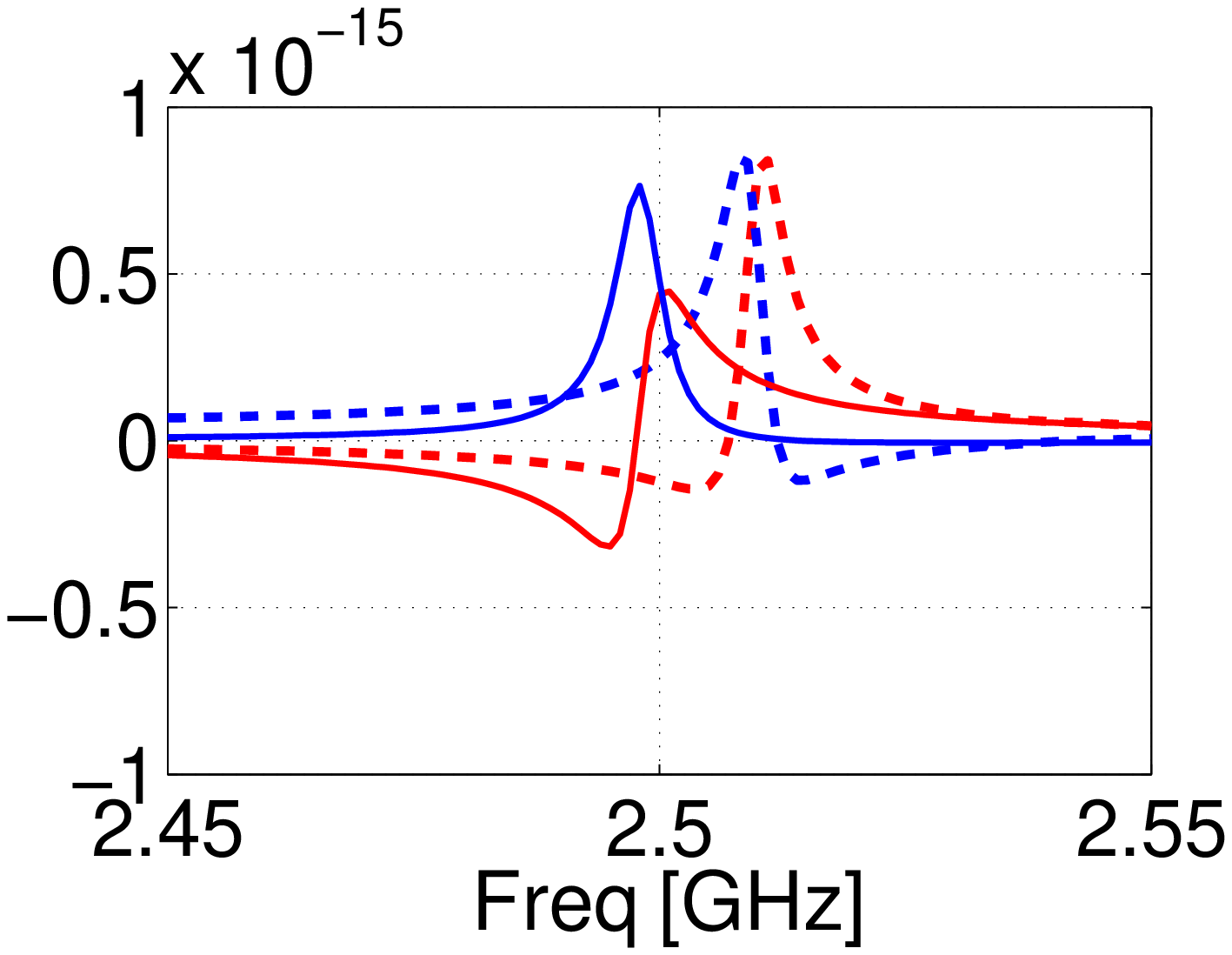}
        \label{fig:j_4}
    }
    \caption{Comparison of analytical and simulated (electro-magnetic and magneto-electric) polarizabilities of the moving-chiral particle with $l=6.9\,\rm{mm}$ and $l'=0.6\,\rm{mm}$. Solid lines are the analytical polarizabilities and dashed lines are the simulated polarizabilities. Red and blue colors represent the real and imaginary parts of the polarizability, respectively.}
    \label{fig:j}
\end{figure}

In this example we have assumed that the long parts of the wires ($l$) are large compared to $l'$, but small compared to the wavelength. Hence, we need to decrease the resonance frequency used in the example of the Tellegen-omega particle (10 GHz) to be able to increase the value of $l$, so that the wavelength would be still larger than the size of particle. Therefore, 2.5 GHz can be a reasonable choice, corresponding the wavelength 120 mm. The ferrite material is the same as above (the same relative permittivity, the saturation magnetization and damping factor). For having the resonance at 2.5 GHz, the applied bias field is  892.5 $\rm{Oe}$. We assume $l'=0.6$ mm, and $l=6.9$ mm. The analytical results for the particle are compared with the simulated ones in Fig.~\ref{fig:i} and Fig.~\ref{fig:j}.
Fig.~\ref{fig:i} shows the electric and the magnetic polarizabilities, and Fig.~\ref{fig:j} shows the electro-magnetic and the magneto-electric polarizabilities. As it is clear in the figures, there is a very small difference in the resonance frequency between the analytical and the simulated results. That can be because of inaccuracy in calculating the applied bias field for having the resonance at 2.5 GHz. However, it is thought that the main source of that shift is in inaccuracies of determining the reactive part of the input impedances of the wire elements. The analytical polarizabilities: $\aeeo$, $\aeer$, $\ammo$, and $\ammr$ are nicely matched in value and behavior with the simulated results. Let us next discuss the behavior of the electro-magnetic and magneto-electric polarizabilities, where we observe differences between analytical and simulated results.

\begin{figure}[h!]
    \centering
    \subfigure[$\aemo$]
    {
        \includegraphics[width=4cm]{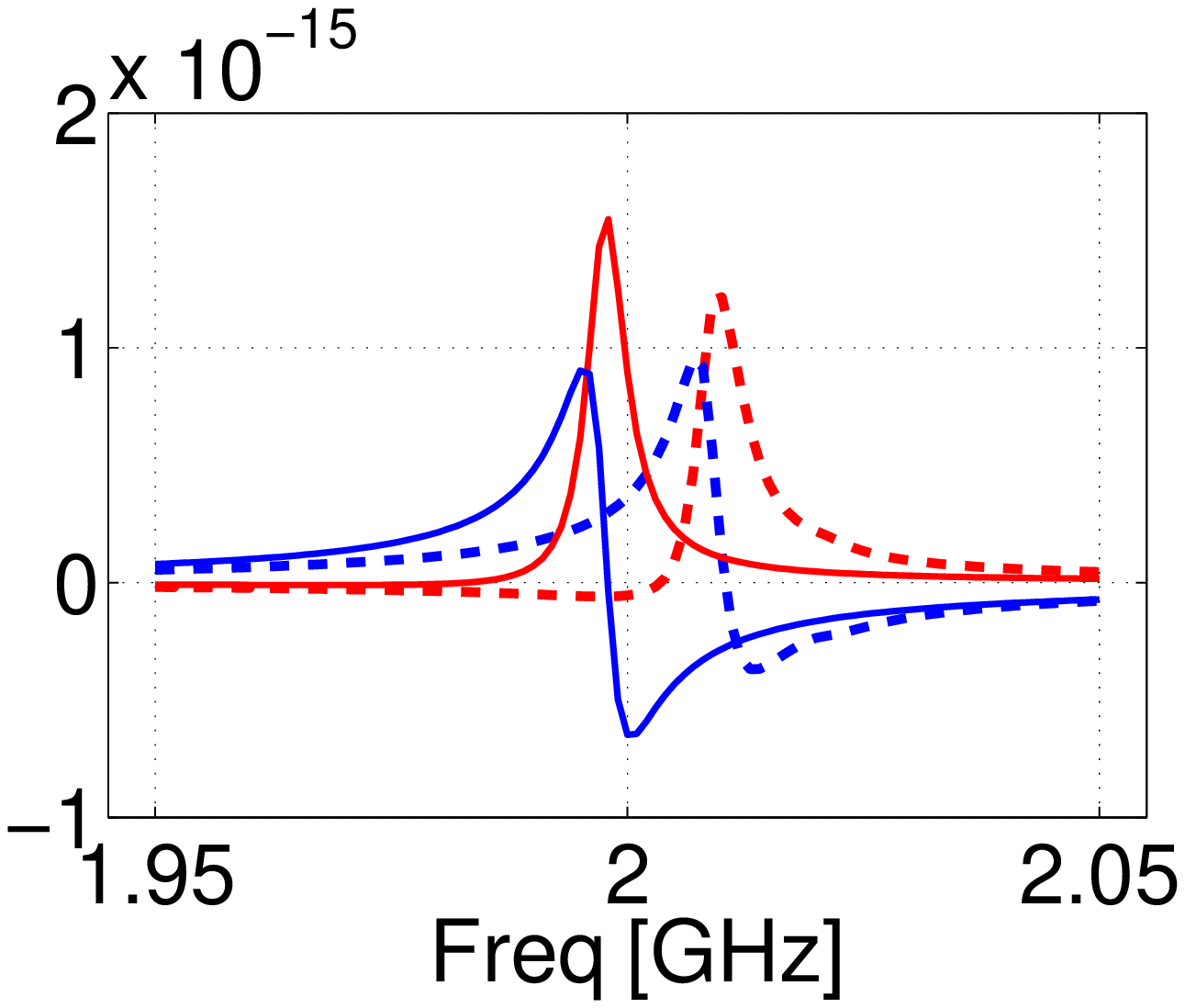}
        \label{fig:10002_1}
    }
    \subfigure[$\ameo$]
    {
        \includegraphics[width=4cm]{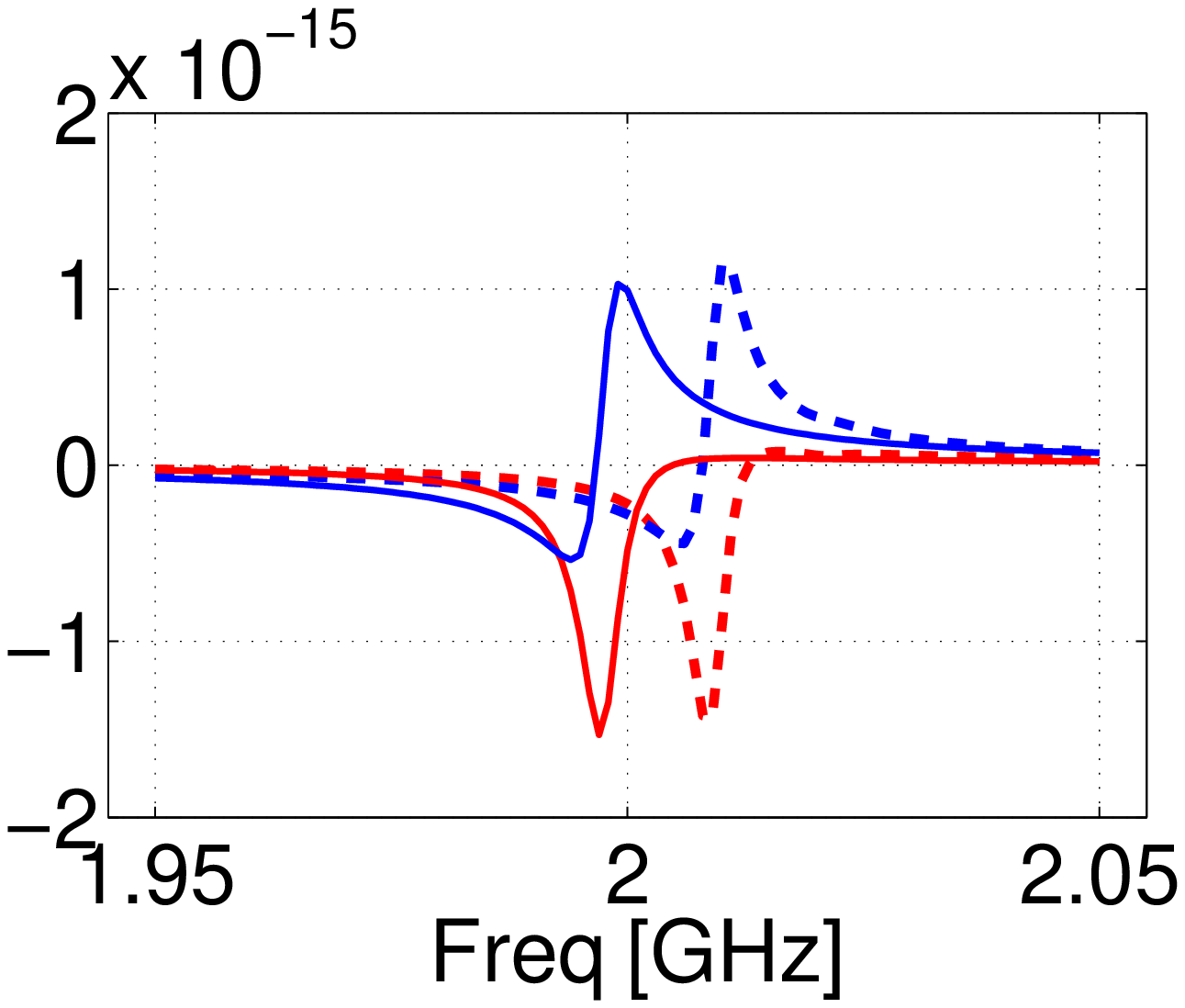}
        \label{fig:10002_2}
   }
     \\
    \subfigure[$\aemr$]
    {
        \includegraphics[width=4cm]{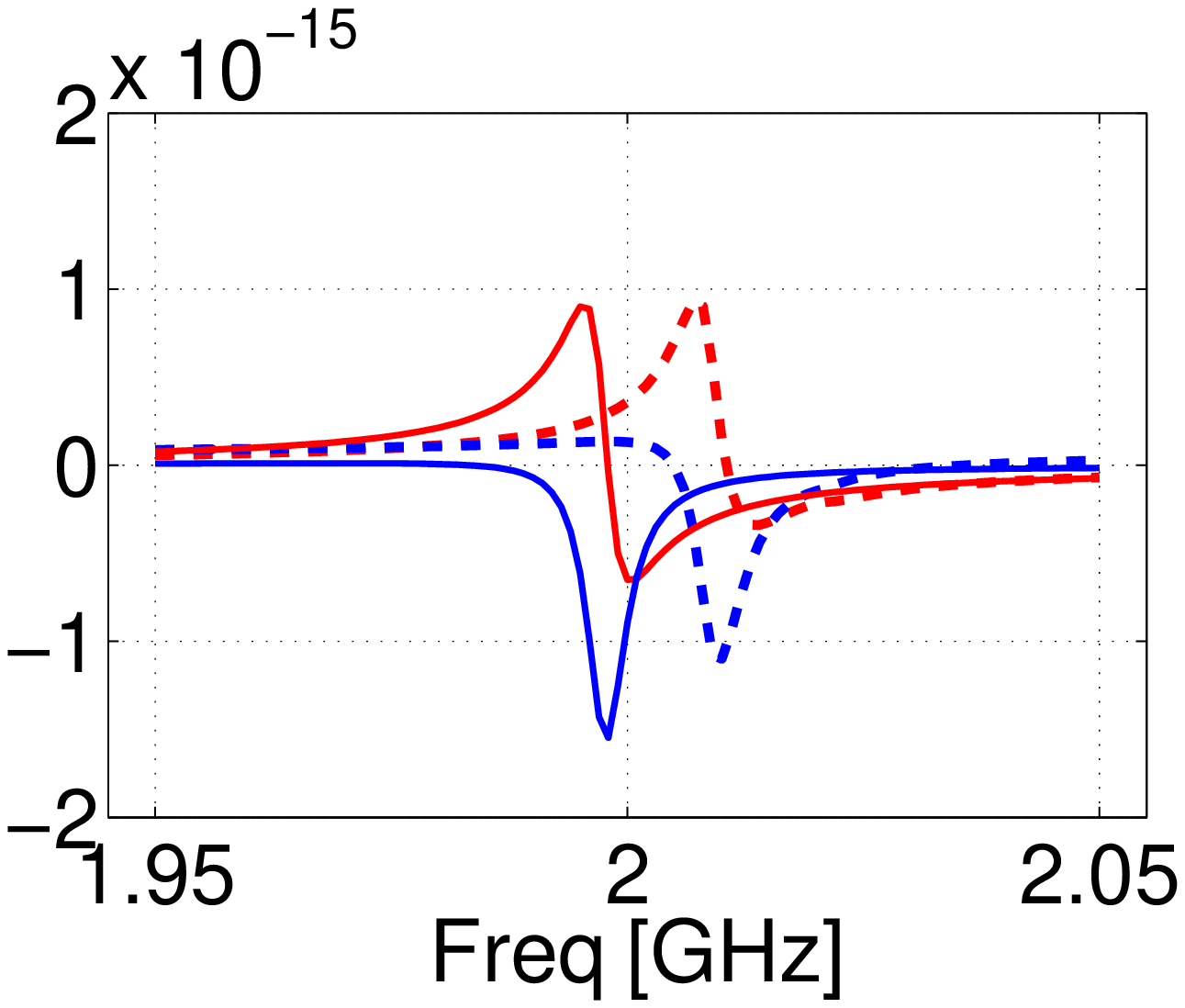}
        \label{fig:10002_3}
    }
    \subfigure[$\amer$]
    {
        \includegraphics[width=4cm]{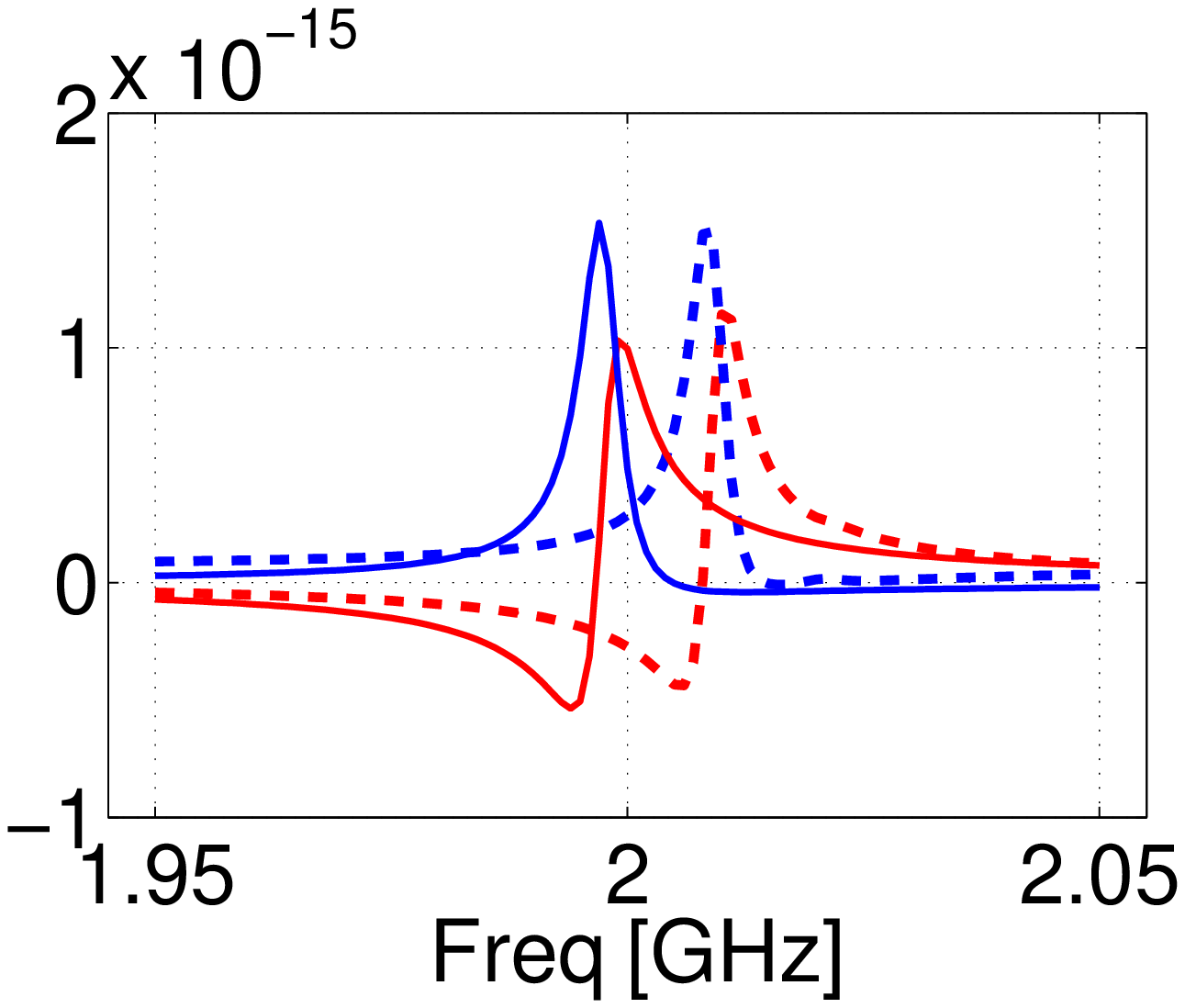}
        \label{fig:10002_4}
    }
    \caption{Comparison of analytical and simulated (electro-magnetic and magneto-electric) polarizabilities of the moving-chiral particle with $l=9\,\rm{mm}$ and $l'=1.5\,\rm{mm}$. Solid lines are the analytical polarizabilities and dashed lines are the simulated polarizabilities. Red and blue colors represent the real and imaginary parts of the polarizability, respectively.}
    \label{fig:10002}
\end{figure}

If we focus only on the analytical results, we find that the polarizabilities $\aemo$ and $\ameo$ or the polarizabilities $\aemr$ and $\amer$ are not exactly opposite to each other in the whole frequency range from 2.45 GHz to 2.55 GHz. For a pure moving-chiral particle, we should have $\aemo=-\ameo$ (chiral property) and $\aemr=-\amer$ (moving property). In fact, with the structure of our particle, there are also couplings of the other two types: Tellegen and omega, in addition to the moving and chiral ones. The fundamental reason for existing Tellegen and omega coupling effects are the central (short) parts of the wires close to the ferrite inclusion. Therefore, we observe mixing of all four coupling phenomena and not a pure moving-chiral particle. Looking at the simulated results for electro-magnetic and magneto-electric polarizabilities, we see that this is more pronounced in the  numerical results. We think that in addition to this topology effect, the long arms of the wires are very close to the ferrite inclusion and they can create parasitic effects in the ferrite sphere excitation (due to strong non-uniformity of the fields). In theory we have not taken into account  these parasitic effects, and therefore the differences of the numerical electro-magnetic and magneto-electric polarizabilities and  the theoretical ones 
are not negligible. Hence, we should find a way to reduce the impact of the long arms of the wires, and one possible solution is moving them further away from the ferrite sphere by increasing the length of the central part of the wires. We have studied an example case with $l'=1.5$ mm. As mentioned before, $l$ is assumed to be much larger than $l'$. To realize this, we suppose that $l$ is six times larger than $l'$, which gives $l=9$ mm. Because the size of the particle should be much smaller than the wavelength, we replace 2.5 GHz with 2 GHz as the resonance frequency. The corresponding applied bias field will be 714 Oe. Fig.~\ref{fig:10002} presents the real and imaginary parts of the electro-magnetic and the magneto-electric polarizabilities obtained analytically and numerically for this example.

As it is seen, now there is fairly good agreement between the analytical and simulated results. Still because of inevitable inaccuracies, a little difference exists in the resonance frequencies. The parasitic effects have been mainly removed but, naturally, the Tellegen and omega coupling effects remain significant. 

\section{Conclusion}
In this work, we have applied the antenna theory concepts (for electrically small short-circuit wire antennas) and the knowledge about electromagnetic properties of ferrite materials to derive analytically the electric, magnetic, electro-magnetic and magneto-electric polarizabilities of two artificial nonreciprocal bianisotropic particles, named Tellegen-omega and moving-chiral. In these particles, the values of all the polarizabilities can be engineered by properly choosing the topology and sizes of the metal parts and the parameters of the magnetized ferrite sphere. We have studied excitations of these particles in incident uniform electric fields in the plane of the particles and found  the electric and magneto-electric polarizabilities. Studies of excitations by incident uniform magnetic fields allowed us to derive the magnetic and electro-magnetic polarizabilities. Subsequently, we have compared the analytical results with the simulated ones. For the Tellegen-omega particle, which has a simpler topology, the analytical polarizabilities showed very good agreement with the numerical polarizabilities. For the moving-chiral particle, because of the complex shape of the metal parts, parasitic effects due to field inhomogeneities are more significant, and initially the analytical results did not agree well with the simulated results. However, we have shown that optimization of the particle dimensions allows us to reduce the parasitic interactions, and we have observed that the analytical and simulated results are almost identical for the optimized particle. Comparison of the electro-magnetic and magneto-electric polarizabilities for the particle indicated that the particle exhibits  also some Tellegen and omega coupling effects. 

In the future, we hope to confirm the analytical and numerical polarizabilities also experimentally. By developing waveguide measurement techniques and placing such particles inside a waveguide with different orientations, it is thought that it can be possible to find all the relevant components of the particle polarizabilities. The other important research problem is to find possible means to control different magneto-electric coefficients independently, so that any desired (physically possible)  values of the parameters would be at reach. As we see from the results of this paper, the example ``moving''  particle shows also some degree of Tellegen and omega coupling, which may be detrimental for some potential applications. We hope that this paper is a significant step towards future use of  nonreciprocal bianisotropic particles in microwave devices, such as nonreciprocal perfect absorbers or nonreciprocal thin-sheet isolators.

\end{document}